\UseRawInputEncoding
\documentclass[
aps,
prx,
twocolumn,
superscriptaddress,
nofootinbib,
amsmath,amssymb,
floatfix]{revtex4-2}

\usepackage{graphicx}
\usepackage{dcolumn}
\usepackage{bm}
\usepackage[dvipsnames]{xcolor}
\usepackage[%
    colorlinks=true,
    pdfborder={0 0 0},
    linkcolor= black,
    citecolor= black
]{hyperref}
\usepackage{etoolbox}
\usepackage[flushleft]{threeparttable}
\usepackage{ulem}

\newcommand{\abthree}[1]{\textcolor{black}{#1}}
\newcommand{\qstate}[1]{$|{#1}\rangle$}
\DeclareMathOperator{\tr}{tr}

\makeatletter

\makeatletter

\makeatletter
\def\@email#1#2{%
 \endgroup
 \patchcmd{\titleblock@produce}
  {\frontmatter@RRAPformat}
  {\frontmatter@RRAPformat{\produce@RRAP{*#1\href{mailto:#2}{#2}}}\frontmatter@RRAPformat}
  {}{}
}%
\makeatother
\begin{document}

    \title{Readout-induced leakage of the fluxonium qubit}

    \author{Aayam Bista}
    \affiliation{%
    Department of Physics, University of Illinois Urbana-Champaign, Urbana, IL 61801}
    \author{Matthew Thibodeau}
    \affiliation{%
    Department of Physics, University of Illinois Urbana-Champaign, Urbana, IL 61801}
    \author{Ke Nie}
    \affiliation{%
    Department of Physics, University of Illinois Urbana-Champaign, Urbana, IL 61801}
    \author{Kaicheung Chow}
    \affiliation{Holonyak Micro $\&$ Nanotechnology Lab, University of Illinois Urbana-Champaign, Urbana, IL 61801}
    \author{Bryan K Clark}
    \affiliation{%
    Department of Physics, University of Illinois Urbana-Champaign, Urbana, IL 61801}
    \author{Angela Kou} 
    \affiliation{%
    Department of Physics, University of Illinois Urbana-Champaign, Urbana, IL 61801}
    \affiliation{Holonyak Micro $\&$ Nanotechnology Lab, University of Illinois Urbana-Champaign, Urbana, IL 61801}
    \affiliation{Materials Research Laboratory,
    University of Illinois at Urbana-Champaign, Urbana, IL 61801}

    \date{\today}

    \begin{abstract}
    Dispersive readout is widely used to perform high-fidelity measurement of superconducting qubits. Much work has been focused on the qubit readout fidelity, which depends on the achievable signal-to-noise ratio and the qubit relaxation time. As groups have pushed to increase readout fidelity by increasing readout photon number, dispersive readout has been shown to strongly affect the post-measurement qubit state. Such effects hinder the effectiveness of quantum error correction, which requires measurements that both have high readout fidelity and are quantum non-demolition (QND). Here, we experimentally investigate non-QND effects in the fluxonium. We map out the state evolution of fluxonium qubits in the presence of resonator photons and observe that these photons induce transitions in the fluxonium both within and outside the qubit subspace. We numerically model our system and find that transitions to higher-excited states and coupling to an external spurious mode are necessary to explain observed non-QND effects.
    \end{abstract}

    \date{\today}

    \maketitle
    
    \section{Introduction}
    A particular strength of superconducting qubit platforms is the ability to perform dispersive readout of the qubit where the qubit state is reflected in the frequency of a coupled readout resonator \cite{Gambetta_Qubit-photon}. 
    Multiple groups have used dispersive readout to achieve very high signal-to-noise ratios (SNRs) for readout, leading to readout fidelities, which characterize the probability that the readout outcome reflects the qubit state, of $>99\%$ \cite{walter_rapid_2017,Sunada_FastReadout}.
    The main method for improving readout fidelities has been to increase the number of photons used for reading out the resonator. 
    It has been observed by multiple groups, however, that dispersive readout using increased photon numbers is not quantum non-demolition (QND), i.e. the post measurement state is changed from the pre-measurement state by the measurement \cite{Khezri_MISTWithin,Sank_MISTBeyond,Vool_Non-Poissonian,Kou_SimultaneousMonitoring,Ficheux_FastLogic,Gusenkova_QND}. 
    This presents a significant challenge for the implementation of quantum error correction, which requires that measurements have both high readout fidelities and are QND. 
    For quantum error correction, it is the QND measurement fidelity $\mathcal{F}^\mathrm{QND}=\frac{1}{2}[P(g,0|g)+P(e,1|e)]$, where $P(q,q_m|q)$ is the probability that a qubit initialized in the state \qstate{q} is measured to be in \qstate{q} with corresponding measurement outcome $q_m$ and remains in \qstate{q} post-measurement, that must be optimized \cite{hazra_benchmarking_2025}. Knowledge of the non-QND effects of dispersive readout and their origins is necessary to determine how best to maximize $\mathcal{F}^\mathrm{QND}$.
    
    Observed non-QND effects have manifested differently in different types of superconducting qubits. In transmons, non-QND effects have been shown to arise from leakage to higher levels of the transmon potential \cite{Dumas_TransmonIonization,Khezri_MISTWithin,Sank_MISTBeyond}. 
    In the fluxonium, on other hand, non-QND effects have been observed in the form of decreased qubit relaxation time when the resonator is populated with only a few photons \cite{Vool_Non-Poissonian,Kou_SimultaneousMonitoring,Somoroff_ms_coherence,Ding_HighFidelity}. 
    This negatively impacts the QND measurement fidelity in two ways: it decreases the readout fidelity, which is exponentially sensitive to the relaxation time \cite{Gambetta_Protocols,Krantz_QuantumEngineer} and decreases the QNDness.
    
    The origin of the non-QND effects in fluxonium remains unclear, and without a model for them it is difficult to optimize device construction and measurement protocols.
    Recent theoretical work has conjectured that the observed non-QND effects in the fluxonium are due to measurement-induced transitions to higher-energy fluxonium states \cite{Nesterov_MIST} or higher-frequency parasitic modes residing in the fluxonium superinductance \cite{singh_impact_2024}.
    Experiments on fluxonium readout have mainly reported on achieved readout fidelity in the presence of non-QND effects with different devices exhibiting non-QND effects at drastically different photon numbers \cite{Vool_Non-Poissonian,Kou_SimultaneousMonitoring,Somoroff_ms_coherence,Ding_HighFidelity,Gusenkova_QND}. It is crucial, however, that we understand the origins of these non-QND effects in order to improve fluxonium QND measurement fidelity and make progress on building fluxonium-based processors.
    
    Here we experimentally investigate the effects of resonator photon number on the fluxonium qubit state. 
    Our observations demonstrate that decreases in relaxation times of the fluxonium in the presence of readout photons are due to transitions both within and outside of the qubit subspace. 
    We find both a monotonic decrease of QNDness with respect to the resonator population alongside additional non-monotonic dips of the relaxation time at particular population values.
    We numerically model the observed non-QND effects in our data and find that the inclusion of a spurious mode into a driven fluxonium-resonator Hamiltonian is necessary to model observed qubit transition probabilities.
    By varying the applied external flux to change both the transition selection rules and fluxonium frequency, we identify the importance of defect-based two-level systems in causing observed non-QND effects.
    We thus find that both resonator-driven transitions to higher-energy fluxonium states and resonator \abthree{photon}-induced coupling to spurious modes are responsible for non-QND behavior in the fluxonium.
    We finally discuss implications of this model for improving and controlling dispersive readout in the fluxonium.
    
    \begin{figure}
    \centering
        \includegraphics{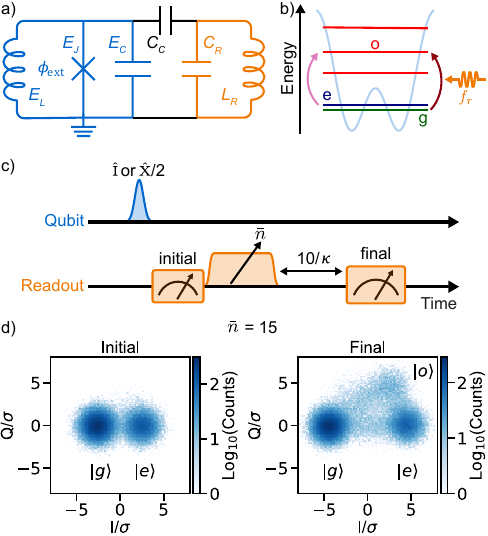}
        \caption{\label{fig:1} 
            (a) Circuit schematic of the device. A fluxonium (blue) with inductive energy $E_L$, Josephson energy $E_J$, and capacitive energy $E_C$ is capacitively coupled (red) with mutual capacitance $C_C$ to a readout resonator (green) with capacitance $C_R$ and inductance $L_R$. The flux applied to the fluxonium is denoted by $\phi_\mathrm{ext}$  
            (b) Energy level diagram of the fluxonium. 
            A periodic drive at the resonator frequency, $f_r=\omega_r/2\pi$, can induce transitions from \qstate{g} or \qstate{e} to levels outside the qubit subspace denoted by \qstate{o}. 
            (c) Pulse sequence for investigating the effect of readout photons on the fluxonium state. An $\hat{\mathrm{I}}$ or $\hat{\mathrm{X}}$/2 operation is first applied to the fluxonium. 
            We then perform an initial  projective measurement of the fluxonium state followed by a varying drive to introduce $\bar{n}$ photons into the resonator. 
            A final stronger and longer projective measurement is then performed on the fluxonium. 
            (d) Readout histograms of the fluxonium qubit before and after 15 photons have been introduced in the resonator. 
            The state corresponding to each distribution is labeled in black.}
    \end{figure}

    \section{Probing the fluxonium in the presence of resonator photons}
    We investigated the behavior of multiple fluxonium devices (parameters in Table I) coupled to driven resonators (Fig.~\ref{fig:1}(a),(b)) by monitoring their state evolutions in the presence of resonator photons (Fig.~\ref{fig:1}(c)) \cite{Sank_MISTBeyond}. We will focus here on Device A before discussing the other devices in Section VI.
    \abthree{In our experiment, we first either use the $\hat{\mathrm{X}}/2$ operation followed by the first measurement, to start in an equal mixture of \qstate{g} and \qstate{e} or apply no pulse and start in the thermal equilibrium population of \qstate{g} and \qstate{e}, to ensure statistically significant initial counts in both states.}
    An initial measurement, using approximately three readout photons and applied for 5~$\mu$s, is then performed to discriminate between the \qstate{g} and \qstate{e} states. 
    Here, we chose a low readout power to mitigate non-QND effects and excitations out of the qubit subspace.  
    Immediately after the initial measurement, we applied a drive on the resonator to mimic the effect of readout photons in the resonator during a measurement pulse.
    We will refer to this drive on the resonator between measurements as the proxy readout drive.
    No measurement result was recorded during the proxy readout drive.  
    After allowing the photons to evacuate the resonator during a ringdown period of 10/$\kappa$, where $\kappa$ is the linewidth of the resonator, we performed a final measurement.
    Here we used a longer (10~$\mu$s) and higher power pulse (readout photon number $\approx$ 4) for the final measurement in order to discriminate between \qstate{g}, \qstate{e} and higher qubit states.
    \abthree{All pulses applied on the resonator are simple square pulses with no additional shaping of the rising and falling edges.}
    \abthree{The effects of pulse shape and rate of photon population change are discussed in section XV of the supplement.}

    We repeated the pulse sequence $10^5$ times for each resonator photon number in order to build statistics on the behavior of the fluxonium qubit during the application of the proxy readout drive. 
    Between the end of a measurement run and the beginning of the next run, we inserted a delay of five times $T_1$ to allow for the qubit to relax back to thermal equilibrium.
    Measurements of the transition times between different fluxonium states were performed to check that this delay time was indeed long enough for the qubit to reach equilibrium.
    Figure~\ref{fig:1}(d) shows readout histograms for the initial and final measurements when 15 photons were introduced by the proxy readout. 
    We fitted these histograms to Gaussian distributions to determine state assignment criteria.
    We set the threshold for state discrimination at the center between the Gaussian distributions corresponding to \qstate{g} and \qstate{e} \cite{Supplement}.
    The readout fidelity is determined by the SNR and the fluxonium $T_1$ \cite{Gambetta_Protocols}. 
    The initial measurement has readout fidelity $\approx$ 95\% and the final measurement has readout fidelity $\approx$ 88\%. 
    The decreased readout fidelity of the second measurement is due to the longer measurement time.
    
    In the initial measurement, we observe population in only the \qstate{g}  and \qstate{e} states. 
    After populating the resonator with 15 photons, the final measurement shows a decrease in the population of the excited state and the appearance of population outside of the qubit manifold. 
    We do not observe the presence of this external population when the resonator is populated with zero photons.
    We performed separate calibration measurements explicitly preparing the fluxonium in the \qstate{f}, \qstate{h} and \qstate{i} states to identify the resonator response when the qubit is outside of the qubit subspace \cite{Supplement}. 
    We found the response for these three states to be indistinguishable from each other but distinguishable from the qubit manifold and consistent with the peak of the Gaussian distribution observed in the right panel of Fig.~\ref{fig:1}(c). 
    We thus denote this third Gaussian distribution as the other (\qstate{o}) state. 
    The presence of this third Gaussian distribution indicates that the photon population in the readout resonator induces leakage out of the qubit subspace to a higher-energy fluxonium state.

    \begin{figure}
    \centering
        \includegraphics{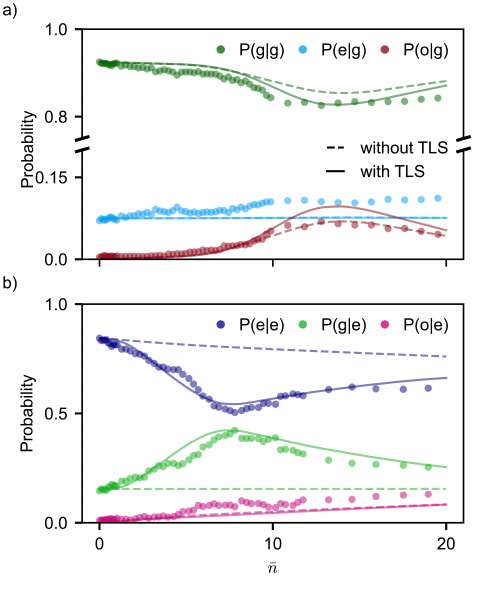}
        \caption{\label{fig:2} 
            Observed (dots) and simulated (lines) probabilities $P(f_0 | i_0)$ of measuring final state \qstate{f_0}, where $f_0 = g$, $e$, or $o$, when starting in initial state \qstate{i_0} for both (a) $i_0 = g$ and (b) $i_0 = e$ as a function of the steady-state photon number $\bar{n}$ in the resonator. 
            The simulations are performed at $\phi_\mathrm{ext}$ = 0.500196. 
            Error bars are smaller than markers in the data.
            \abthree{The dashed (solid) lines show the results of the numerical time-domain simulations of the coupled fluxonium-resonator system in the presence of a resonator drive (with a TLS coupled to the qubit).}}
    \end{figure}
    \section{Photon dependence of induced transitions}

    We repeated the experiment for different numbers of photons introduced by the proxy readout drive to map out the dependence of the observed change in qubit state populations on photon number. 
    The counts of different initial and final states are used to calculate the probability $P(f_0|i_0)$ of ending up in the final state \qstate{f_0}, given that the initial state was \qstate{i_0}. 
    We will use the probability of $P(i_0|i_0)$ as a measure of the QNDness in the presence of the proxy readout drive. 
    Since there is a time delay between the initial and final measurements, both the fluxonium $T_1$ and the number of photons in the resonator will affect $P(i_0|i_0)$. 
    Here we keep the time between the initial and final measurements fixed to focus solely on the effect of resonator photons in inducing non-QND effects.
    \abthree{When there are zero photons in the resonator, we attain the maximal QNDness possible for our device setup, which includes the effects of fluxonium dissipation and of any non-QNDness in the initial and final measurement pulses.}
    The results for zero to 20 photons in the resonator with initial state \qstate{g} (\qstate{e}) \abthree{at $\phi_\mathrm{ext}=0.5$ are shown in Fig.~\ref{fig:2}(a) (Fig.~\ref{fig:2}(b)). The qubit frequency $\omega_q/2\pi =401.5$ MHz at this flux.} 
    The data is corrected for the expected state assignment error based on the signal-to-noise ratio (SNR) obtained from Gaussian fits to the readout histograms \cite{Gambetta_Protocols,Krantz_QuantumEngineer,Supplement}. 

    While we observe non-QND effects arising when starting out in both \qstate{g} and \qstate{e}, the observed non-QNDness is significantly weaker when starting in \qstate{g}. 
    The probability decrease of remaining in \qstate{g} is observed with a corresponding rise in the probability of induced transitions to \qstate{e} and \qstate{o}, which indicates resonator photon-induced leakage outside of the qubit subspace. 
    The non-QNDness of resonator photons when starting in \qstate{e} is significantly worse than when starting in \qstate{g}. 
    The probability of staying in \qstate{e} already significantly decreases to 0.5 when only seven photons are introduced  into the resonator. 
    The probability unexpectedly increases, however, between $\bar{n} \approx$ 7 and $\bar{n} \approx$ 12, which suggests resonant behavior induced by the resonator photons centered at $\bar{n} \approx$ 7.
    \section{Modeling the oberved fluxonium behavior}
    To quantitatively understand the origins of the observed phenomena, we perform numerical time-domain simulations of the coupled fluxonium-resonator system in the presence of a resonator drive:
    
    \begin{align} \label{drive}
    \hat{H}_\mathrm{driven} = \hat{H}_f+\omega \hat{a}^\dag_r \hat{a}_r-i g\hat{n}(\hat{a}_r-\hat{a}^\dag_r) 
    \notag
    \\ -i \epsilon \cos(\Omega t)(\hat{a}_r-\hat{a}^\dag_r),
    \end{align}
    where $\hat{H}_f$ is the fluxonium Hamiltonian \cite{Vladimir_Fluxonium}, $\hat{H}_r$ is the resonator Hamiltonian, $g$ is the coupling between the resonator and the fluxonium, $\hat{n}$ represents the number of excess Cooper pairs on the fluxonium capacitance, $\hat{a}_r$ and $\hat{a}^\dag_r$ are the resonator ladder operators, and $\epsilon$ and $\Omega$ is the strength and frequency of the resonator drive. 
    A similar Hamiltonian has been used to describe non-QND effects in the transmon \cite{Sank_MISTBeyond,Khezri_MISTWithin,Dumas_TransmonIonization} and in recent theoretical work on the fluxonium \cite{Nesterov_MIST}. 
    
    We truncate the Hamiltonian to a finite Hilbert space and use Lindbladian dynamics to simulate the time evolution of this Hamiltonian combined with resonator dissipation \cite{Supplement}. 

    The predicted behavior of the transition probabilities is shown in Fig.~\ref{fig:2} using dashed lines; fluxonium and resonator parameters used in the model are extracted from independent spectroscopy measurements \cite{Supplement}.
    This model predicts that the resonator photons induce transitions to the \qstate{i} state of the fluxonium from \qstate{e}. 
    At exactly $\phi_\mathrm{ext}=0.5$, transitions from \qstate{g} are forbidden by parity.
    The decrease in QNDness observed at higher photon number when starting in \qstate{g} suggests that we are not exactly at $\phi_\mathrm{ext}=0.5$ and we find that changing the flux by an experimentally-undetectable offset of $0.000196 \phi_0$ fits our data well. 
    \abthree{The prediction at exactly $\phi_\mathrm{ext}=0.5$ is provided in section XIII of the supplementary material \cite{Supplement}}.
    We see that the evolution of the \qstate{g} state can be reasonably modeled by the driven fluxonium-resonator system though we do see that there is disagreement between the theoretical model for the population in the excited states at larger photon numbers when compared with the measured data.
    We hypothesize that this discrepancy may be due to qubit decay from higher excited states into the \qstate{e} state, which is not included in our model.
    We note that we subtract a constant probability offset from the simulated data to compensate for fluxonium dissipation not being included in our simulations.

    Multiple features of the behavior of the \qstate{e} population are, however, not captured by this model. 
    While the numerics show some decrease in the probability of remaining in \qstate{e} with increasing photon number, the measured decrease is significantly faster. 
    Neither this drastic decrease in the \qstate{e} state nor the resonance-like behavior at $\bar{n}\approx 7$ is captured by introducing this drive term. 
    This suggests that additional effects beyond resonator-induced transitions to higher-energy fluxonium states are at play in the fluxonium.

    We conjecture that a second effect of the resonator population is to bring the fluxonium \qstate{g}-\qstate{e} transition on resonance with an external lossy mode via the ac-Stark shift \cite{Thorbeck_Readout,sivak_QEC,Connolly_Full}. 
    If the gap of this mode $\mu$ is only slightly greater than the qubit frequency $\omega_q$, a small resonator population can cause significant hybridization of the \qstate{e, 0_{\mu}} and \qstate{g, 1_{\mu}} states. 
    
    We model this additional mode as a two-level system with frequency $\Delta_{\mathrm{TLS}}$ capacitively coupled to the fluxonium with coupling $g_{\mathrm{TLS}}$:
    
    \begin{equation} \label{TLS}
    \hat{H}_\mathrm{TLS}=\hat{H}_\mathrm{driven}+\Delta_{\mathrm{TLS}}\hat{Z}_\mathrm{TLS}+g_{\mathrm{TLS}}\hat{X}_\mathrm{TLS}\hat{n},
    \end{equation}
    
    \noindent where $\hat{Z}_\mathrm{TLS}$ and $\hat{X}_\mathrm{TLS}$ are the Pauli $Z$ and $X$ operators defined in the TLS basis.
    Fits to our data using Eq.~\ref{TLS} are indicated using solid lines in Fig.~2(a) and (b) with $\Delta_{\mathrm{TLS}}/2\pi$~=~411~MHz and $g_{\mathrm{TLS}}/2\pi~=~$1.3~MHz.
    The transitions caused by the spurious mode strongly depend on its thermalizing bath \cite{Supplement}.
    Here, our modeling assumes that the spurious mode is \abthree{completely} initialized in its ground state and predicts that coupling to the mode mainly generates transitions from the excited state.
    \abthree{We discuss the effects of the temperature of the TLS in our model in section XVI of the supplementary material \cite{Supplement}.}
    We see that the introduction of this spurious mode captures well the measured data when the qubit is initialized in \qstate{e}.

    Qubit spectroscopy performed as a function of flux in this frequency range suggests the presence of a TLS mode at $\sim$409~MHz, which is close to the TLS frequency found in our model. \cite{Supplement}
    \abthree{Our model assumes that the ac-Stark shift causes direct resonance between the qubit and the TLS.
    Alternatively, a multi-photon resonance between the qubit and a higher frequency spurious mode would result in similar behavior, which we discuss in section XVII of the supplement \cite{Supplement}.}
    We additionally observe that the non-QND behavior changes for different cooldowns \cite{Supplement}, further suggesting that materials defects are responsible for the observed dips in QND-ness since one would expect array modes or sampleholder modes to stay constant through different cooldowns.
 
    \begin{figure}[h!]
    \centering
        \includegraphics{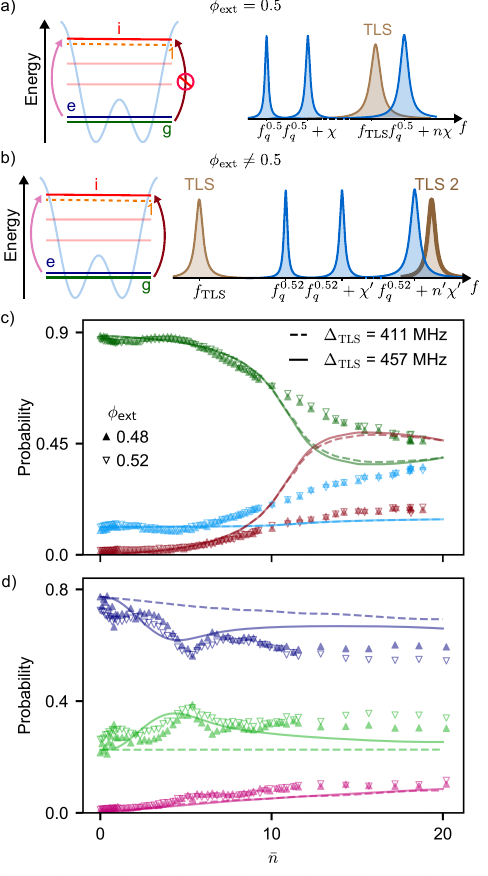}
        \caption{\label{fig:3} 
            (a) Effect of resonator photons at $\phi_\mathrm{ext}=0.5$.  
            At $\phi_\mathrm{ext}=0.5$, parity forbids transitions from \qstate{g} to the state closest to the resonator frequency \qstate{i}. 
            The driven resonator brings the Stark-shifted fluxonium qubit on resonance with a TLS mode.
            (b) Effect of resonator photons at $\phi_\mathrm{ext}\neq0.5$.  
            Transitions from \qstate{g} to \qstate{i} are now allowed. 
            The fluxonium \qstate{g}-\qstate{e} frequency is far detuned from the TLS that it was coupled to at $\phi_\mathrm{ext}=0.5$, but can be Stark-shifted into resonance with a different TLS.
            (c),(d) The probability of the qubit staying in \qstate{g} (\qstate{e}) or transitioning to \qstate{e} (\qstate{g}) or \qstate{o} as a function of $\bar{n}$ at $\phi_\mathrm{ext} = 0.48$ and 0.52.
            Thin solid lines indicate the fit at $\phi_\mathrm{ext}=0.52$ using the same TLS parameters as at $\phi_\mathrm{ext} = 0.5$.
            \abthree{The colors share the same legend as Fig. 2 and the shared legend for the markers are shown in (c).} 
            Solid lines indicate the fit using $\Delta_{\mathrm{TLS}}/2\pi=457$~MHz and $g_{\mathrm{TLS}}/2\pi=1$~MHz \abthree{while dashed lines show the fit using the same TLS parameters as Fig 2}.}
    \end{figure}
    
    \section{Flux dependence of induced transitions}
    Within the framework of our model, we expect breaking parity selection rules to result in increased transitions from \qstate{g} and changing frequency to change the photon number at which resonances with the TLS occur as shown in Fig. \ref{fig:3}(a-b).
    We may investigate each of these effects by monitoring the fluxonium state evolution as a function of readout photon number away from $\phi_\mathrm{ext}=0.5$.

    Our simulations predict that \qstate{i} is the fluxonium level most likely to participate in resonator-induced transitions.
    While transitions from \qstate{g} to \qstate{i} are forbidden at $\phi_\mathrm{ext}=0.5$, they are readily allowed away from this flux point.
    We thus expected and indeed observed a degradation of QNDness for starting in \qstate{g} but not \qstate{e} when we performed our experiment at $\phi_\mathrm{ext} = 0.48$ \abthree{where $\omega_q/2\pi = 449.4$ MHz} (Fig.~\ref{fig:3}(c)).
    The probability of remaining in \qstate{g} shows a larger decrease at $\phi_\mathrm{ext} = 0.48$ compared to $\phi_\mathrm{ext} = 0.5$ and induced transitions to \qstate{e} and \qstate{o} increase.
    We also observe similar behavior between $\phi_\mathrm{ext} = 0.52$ and $\phi_\mathrm{ext} = 0.48$, as expected for the fluxonium, which is symmetric about $\phi_\mathrm{ext} = 0.5$ . 
    
    The predicted behavior using Eq.~\ref{TLS} is shown in \abthree{dashed} lines in Fig.~\ref{fig:3}(c) and (d). 
    Here all parameters for the model except $\phi_\mathrm{ext}$ are the same as in Fig.~\ref{fig:2}. 
    We note that our theoretical model well-predicts the decrease in QNDness off half-flux for starting in \qstate{g} at low photon numbers but starts to deviate at larger photon numbers.
    Our data shows a rise both in transitions from \qstate{g} to \qstate{e} and \qstate{g} to \qstate{o} while the model expects a much more significant increase in transitions between \qstate{g} and \qstate{o}. 
    Similar to the behavior at $\phi_\mathrm{ext}=0.5$, we suspect that this difference in predicted population versus measured population is due to decay from higher-order states during the time of our measurement that is not included in our simulation.
    Decay from \qstate{i} to \qstate{g} or \qstate{e} during our experiment would lead to increased $P(g|g)$ or $P(e|e)$ values compared to the simulation. 
    \abthree{Based on the $T_1$ measurements detailed in Section VII of the supplement, we estimate that $\sim 60\%$ of the population has been transferred from \qstate{i} to \qstate{e}. We speculate that the additional $\sim 8\%$ population discrepancy between data and simulation arises from either the now-allowed direct dissipation to \qstate{g} or decay via the \qstate{f} or \qstate{h} states.} 

    We observe an additional discrepancy between our data and model prediction using the same TLS frequency as at $\phi_\mathrm{ext} = 0.5$. 
    Rather than moving to much higher photon number, the resonance-like feature in the transition probability moves to lower photon number and changes shape. 
    This is unexpected if the TLS frequency remains the same. 
    When moving $\phi_\mathrm{ext}$ from 0.5 to 0.48, the fluxonium qubit frequency changes by approximately 50 MHz. 
    Since the fluxonium-resonator $\chi/2\pi=0.9$~MHz, photons introduced by the proxy readout would ac-Stark shift the fluxonium away from resonance with any fixed frequency mode that was coupled to the fluxonium at half-flux (Fig. \ref{fig:3}(b) right).
    \abthree{Note that we define $2\chi/2\pi$ to be the frequency separation between the qubit state-dependent frequencies of the resonator.}
    Thus, we would expect a disappearance of the resonance as described by the simulation (solid line) in Fig.~\ref{fig:3}(d).
    We observed in the measurement, however, that there remains a resonance in the transition probability when starting from the excited state. 
    This resonance has a different linewidth and height, and its minimum has moved to lower photon number when compared with the resonance observed at half-flux.
    We observed this behavior at both $\phi_\mathrm{ext} = 0.48$ and 0.52. 
    
    We hypothesize that a different spurious mode couples to the fluxonium at this higher frequency.
    The total area of the Josephson junctions in the superinductance and weak junction in our fluxonium device is $\approx 94~\mu \mathrm{m}^2$.
    Based on recently measured TLS densities in a fluxonium of 0.3-0.45 $\mathrm{GHz}^{-1}\mu \mathrm{m}^{-2}$ (over a range of 0.1-0.4 GHz) \cite{zhuangTLSspec}, we estimate that a strongly-coupled TLS should be present every $\sim30$~MHz in our device.
    The thick solid line in Fig. \ref{fig:3} shows the fit to the data with $\Delta_{\mathrm{TLS}}/2\pi=457$~MHz and $g_{\mathrm{TLS}}/2\pi=1$~MHz.
    All other parameters are fixed to be the same as in the fits in Fig. \ref{fig:2}.
    We see that, consistent with our expectation, the TLS frequency needs to shift by close to 50~MHz in order to maintain resonance with the fluxonium away from half-flux.
    \abthree{We note, however, that the data has a larger deviation from our theoretical predictions compared to what we observe at $\phi_\mathrm{ext} = 0.5$.
    We note that our model only includes a single qubit coupled to a fixed-frequency TLS which does not account for multiple TLS's, dissipation in the TLS, or coupling between TLS's which could lead to the observed discrepancy between the data and experiment.}    
    
    \section{Dependence of non-QND effects on resonator parameters}
    
\begin{figure}
    \centering
        \includegraphics{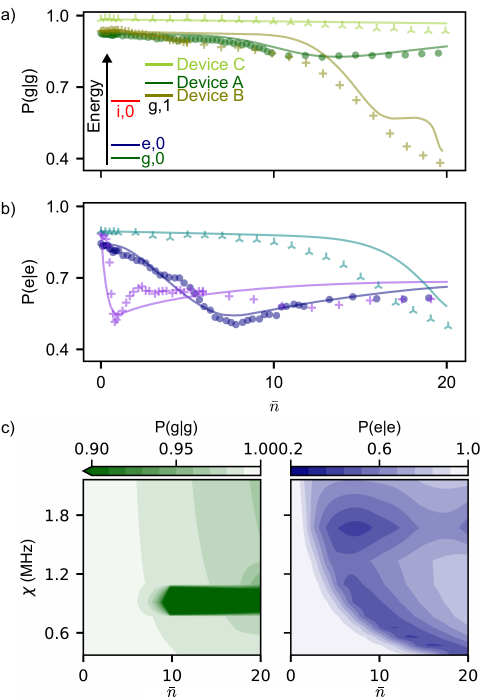}
        \caption{\label{fig:4} 
            (a), (b) Probabilities of staying in the same final state as the initial state for (a) $i_0=g$ and (b) $i_0=e$, for devices with similar fluxonium parameters but varying resonator and coupling parameters. Data from Device A is shown using circles while data from Device B and Device C are shown using cross and tripod markers respectively. Simulations are shown using solid lines. Inset in (a) shows the energy level diagram for the fluxonium and single-photon excitations from the ground state in the different measured devices.
            (c) Simulated probabilities of measuring the same final and initial states for $i_0=g$ (left) and $i_0=e$ (right), for devices with varying resonator frequency to change $\chi$} and resonator photon number. The fluxonium parameters, fluxonium-resonator coupling strength, and TLS parameters are held fixed at the values used to model device C.
    \end{figure}

    We now exploit the qubit-resonator system design freedom enabled by using superconducting circuits and investigate the utility of changing the resonator frequency and coupling strength in mitigating the observed non-QND effects. 
    We repeated the above experiments on two other devices (Device B and C in Table 1) that have similar fluxonium parameters and resonator-fluxonium coupling strength but different resonator frequency (Fig.~\ref{fig:4}(a)-(b)). 
    Due to differences in the proximity of the resonator frequency to the \qstate{i} state, similar coupling strengths ($g$) can yield different dispersive shifts.
    
    In general, reducing $\chi$ by changing the resonator frequency leads to better QNDness in the presence of resonator photons. 
    We observe that Device C, which has the lowest $\chi$ of the three devices, suffers very little from non-QND effects when starting in the \qstate{g} state. 
    Additionally, the probability of leaving the \qstate{e} state decreases the slowest with respect to resonator photon number for Device C. 
    On the other hand, Device B, which has the largest $\chi$, shows non-QNDness independent of the initial state. 
    When the initial state is \qstate{g}, a multi-photon transition between \qstate{g} and a highly excited state of the fluxonium (the 13th excited state as identified by our model) leads to strong non-QND effects starting at ten photons.
    When the initial state is \qstate{e}, the probability of staying in \qstate{e} already significantly drops to below 0.5 in Device B when the resonator contains less than a single photon.
    
    We note that the couplings to TLS's are uncontrolled in our measured devices. 
    As determined by our fits to the data using Eq.~\ref{TLS}, the TLS frequency and strength of coupling differ in the three different devices \cite{Supplement}.
    Our data suggests that materials defects are participating in inducing these non-QND effects. 
    This differing behavior across devices located on the same chip and with similar fluxonium parameters further supports this conclusion. 
    
    In order to identify how controlled design parameters affect the observed non-QND effects, we performed simulations of the QNDness of the readout as a function of resonator photon number and $\chi$ (Fig.~\ref{fig:4}(c)) in the presence of a spurious TLS mode.
    The fluxonium parameters and resonator-fluxonium coupling are kept fixed while the resonator frequency is varied to change $\chi$. 
    The parameters for the simulation are provided in Table ~\ref{table 1}. 
    The TLS frequency and fluxonium-TLS coupling strength are also fixed to 524.5 MHZ and 2.5 MHZ, respectively.
    
    Our simulation shows that resonator photons \abthree{can lead} to coupling to \abthree{lossy} external modes and to states outside of the qubit subspace.
    The QNDness generally shows a very slow decrease with photon number due to weak coupling to the \qstate{h} state when starting in \qstate{g}. 
    This slow decrease, however, is overwhelmed by the significant drop in QNDness when the resonator induces a multi-photon transition between \qstate{g} and a highly-excited state of the fluxonium (dark teal region in left panel of Fig.~\ref{fig:4}(c)). 
    This behavior is similar to previous work on the transmon \cite{Khezri_MISTWithin,Dumas_TransmonIonization}. 
    On the other hand, when starting in the excited state, it is the coupling to a TLS that causes the strongest non-QND effects. 
    We see that the QNDness drops below 0.5 for photon numbers less than 10 for a significant fraction of the phase space. 
    We conclude that designing fluxonium-resonator devices with lower $\chi$ will result in better QNDness, but such devices may still be limited at larger photon numbers by spurious TLS modes.
    Moreover, low-$\chi$ devices will require a longer integration time for high SNR and would likely not be suitable for fast readout.

    \begingroup
    \setlength{\tabcolsep}{7pt}
    \begin{table*}
    \centering
    \caption{Fluxonium and resonator parameters \label{table 1}}
    \begin{tabular}{c  c  c  c  c  c  c  c} 
     Device & $E_J/2\pi$ (GHz) & $E_C/2\pi$ (GHz) & $E_L/2\pi$ (GHz) & $g/2\pi$ (MHz) & 
     $\chi/2\pi$ (MHz) & $\omega_r/2\pi$ (GHz) & $\kappa/2\pi$ (MHz) \\
     \hline
     A  & 2.68  & 1.09 & 0.32 & 203 & 0.9 & 7.440 & 0.6 \\
     
     B  & 2.49  & 1.06 & 0.32 & 215 & 1.8 & 7.047 & 0.6 \\
     
     C  & 2.25  & 1.08 & 0.32 & 250 & 0.1 & 7.826 & 1.4 \\

     Sim  & 2.25  & 1.08 & 0.32 & 250 & 2.16-0.37 & 7.7415-7.8165 & 1.4 \\
    \end{tabular}
    \end{table*}
    \endgroup
    
    \section{Conclusion and outlook}
    We have experimentally investigated the effects of resonator photons on the fluxonium qubit state. 
    Our work substantially increases understanding of fluxonium readout by pinpointing two mechanisms by which dispersive readout is non-QND in the fluxonium: excitation to higher-energy fluxonium states and resonator photon-induced resonance with spurious modes in the device.
    Qubit spectroscopy and the flux dependence of the non-QND effects suggest that the participating spurious modes are likely due to lossy defect-based TLS in the fluxonium circuit.
    Knowledge of these mechanisms enables us to identify fluxonium-resonator parameters to improve QND measurement fidelity and the limits of dispersive readout of the fluxonium.

    We have demonstrated that drive-induced transitions to higher fluxonium levels can be reasonably well-predicted from the Hamiltonian parameters. 
    We note that analogous phenomena in transmons have been well-documented \cite{Dumas_TransmonIonization,Khezri_MISTWithin, Sank_MISTBeyond}.
    The inclusion of decay rates for the different fluxonium states will likely improve the predictive power of our model.
    This predictability demonstrates the utility of time-domain simulations of the driven fluxonium-resonator device as a tool to optimize the design of fluxonium quantum processors. 
    Through prudent choice of fluxonium and resonator parameters, non-QND effects from transitions to higher levels can be minimized. 
    In particular, choosing resonator frequencies that are maximally detuned from certain higher-energy fluxonium levels will result in attenuated transition rates to those levels and hence enable usage of larger photon numbers for increased readout fidelity with minimal non-QND effects.
    In designing such devices, one must also include potential multi-photon resonances which might occur at higher photon numbers.

    The importance of resonator \abthree{photon-induced} resonance with unpredictable TLS's in the fluxonium represents a significantly bigger problem. 
    Since the origin of TLS's in Josephson junctions remains unknown, one cannot design a device to avoid these spurious modes.
    The large total area of the junctions in the fluxonium leads to a large spectral density of TLS's coupling to the fluxonium.
    While the TLS present in the junctions in the superinductance are expected to couple weaker to the fluxonium compared to the TLS present in the weak junction, they have been shown to affect qubit transition rates \cite{zhuangTLSspec}.
    The larger spectral density of TLS's severely limits the number of photons that can be used for readout for a given $\chi$ before the onset of non-QND effects and thus also limits the achievable readout fidelity of a fluxonium device.
    At fixed photon number, lower $\chi$ allows one to avoid resonance via the smaller associated ac-Stark shift, but one must trade this against the reduced readout fidelity per unit time.
    Our simulations indicate that for $\chi \approx 0.5~$MHz, strong non-QND effects will start to arise for $\bar{n}\sim 10$.
    While it was known that unpredictable TLS's represent a bottleneck for development of multi-device quantum processors \cite{zanuz_mitigating_2024}, our results show that such TLS's are also detrimental for qubit readout \cite{sivak_QEC, Thorbeck_Readout}.
    We further expect that next-generation protected qubits that utilize \abthree{Josephson junction arrays} for their inductances such as the 0-$\pi$ qubit \cite{Brooks0Pi, Gyenis0Pi} and the fluxonium molecule \cite{KouFM, KumarProtomon, ThibodeauFFM} will also suffer from similar non-QND measurement effects.
    It is thus of utmost importance to improve Josephson junction fabrication methods or use alternative materials with lower TLS densities.
    
    At the protocol level, we expect our work to spur the development of alternative readout schemes that induce minimal Stark shifts of the fluxonium frequency.
    Additional drives to tune the frequencies of spurious modes during readout may be useful in improving fluxonium readout.
    Longitudinal readout, which couples the qubit state to the imaginary component of the resonator rather than the photon number \cite{didier_fast_2015,touzard_gated_2019,ikonen_qubit_2019}, is a particularly compelling alternative to dispersive readout since it does not require a large $\chi$ and does not induce a large Stark shift on the fluxonium. 

    \section*{Acknowledgments}

    We acknowledge useful discussions with Wolfgang Pfaff, Kyle Serniak, Jeffrey Gertler, Mallika Randeria, Kunal Tiwari, Alexander MacDonald, Alex Chapple, Alexandre Blais, and Alexandru Petrescu. 
    Our traveling-wave parametric amplifier is provided by IBM.
    This research is carried out in part in the Materials Research Lab Central Facilities and the Holonyak Micro and Nanotechnology Lab, University of Illinois. 
    This research is partially supported by the Air Force Office of Scientific Research under award number FA9550-21-1-0327. 
    Research is also sponsored by the Army Research Office and is accomplished under Gates on Advanced Qubits with Superior Performance” (GASP) (Contract No. W911-NF23-10093) and under Grant No. W911NF-23-1-0096. 
    The views and conclusions contained in this document are those of the authors and should not be interpreted as representing the official policies, either expressed or implied, of the Air Force Office, Army Research Office or the U.S. Government.
    
    \bibliography{citation}

\clearpage
\begin{center}
\textbf{\large Supplemental Material}
\end{center}

\setcounter{equation}{0}
\setcounter{figure}{0}
\setcounter{table}{0}
\setcounter{section}{0}
\renewcommand{\theequation}{S\arabic{equation}}
\renewcommand{\thefigure}{S\arabic{figure}}

\section{Device description}

\begin{figure}
    \centering
        \includegraphics{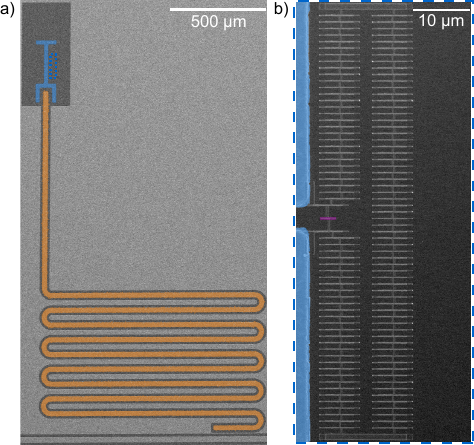}
        \caption{\label{device_image} 
            SEM image of a device fabricated together with the measured devices: (a) SEM image of a full device. False coloring shows the capacitive pads of the fluxonium (blue) \abthree{and} the coplanar waveguide resonator (\abthree{orange}). (b) SEM image of the Josephson junctions that form the fluxonium. False coloring shows the weak junction (magenta). The rest of the junctions form the super-inductance.}
    \end{figure}

    Figure~\ref{device_image}(a) shows an SEM image of a fluxonium-resonator device fabricated along with the measured devices.
    We use a two-dimensional coplanar waveguide (\abthree{orange}) as the readout resonator. 
    We capacitively couple the fluxonium (blue) to the resonator as shown in Figure~\ref{device_image}(a) using a large capacitive pad closer to the resonator to enable stronger coupling. 
    The resonator is capacitively coupled to a two-dimensional coplanar waveguide that acts as a transmission line.
    Control pulses to both the qubit and the resonator were applied through the transmission line. 

    The devices were fabricated on a sapphire substrate.
    The capacitive pads, resonator, transmission line and the ground plane are made of tantalum (Ta).
    The Ta was patterned using optical lithography and wet-etching.
    The Josephson junctions are Al-AlOx-Al junctions made using double-angle electron-beam evaporation after patterning double-layered resist using electron-beam lithography. 
    Figure~\ref{device_image}(b) shows the array of 125 Josephson junctions that forms the shunting inductance and the lone weak junction of the fluxonium (magenta). 
    All the measured devices were fabricated on the same chip and share the same transmission line.
    
    \section{Experimental setup}

    We mixed local oscillator (LO) drives from signal generators (\textit{SignalCore, SC5511A}) with IQ tones from an integrated FPGA system (\textit{Quantum Machines, OPX+}) to create the pulses for readout and proxy readout. 
    We used a separate signal generator (\textit{Rohde \& Schwarz, SGS100A}) with the FPGA to generate the qubit pulses.
    The three signals were combined using a directional coupler before entering a dilution refrigerator (\textit{Oxford Instruments, Triton 500}).
    Twenty dB of attenuation is placed on each of the 4K and 100 mK stages to thermalize the input line. The input line goes through 30 dB of attenuation, a low-pass filter, and an eccosorb filter on the mixing chamber stage before connecting to the sampleholder, which is attached to the mixing chamber.

    The output from the transmission line was sent through a high-pass filter and an eccosorb filter before going through a traveling wave parametric amplifier (TWPA). 
    The signal is further amplified by a high-electron-transistor (HEMT) amplifier (\textit{Low Noise Factory, LNF-LNC4\_16B}) at the 4 K stage and by a room temperature low-noise amplifier (\textit{Low Noise Factory, LNF-LNR4\_14B\_SV}) outside the dilution refrigerator. 
    The signal was then demodulated using the same LO as the readout pulse.
    The demodulated signal is further amplified (\textit{Stanford Research Systems, SR445A}) before being sent to the FPGA for further processing.
    A global coil with a DC current source (\textit{Yokogawa, GS200}) was used to flux bias the devices. 
    
    \section{Device characterization}
    \begin{figure*}
    \centering
        \includegraphics{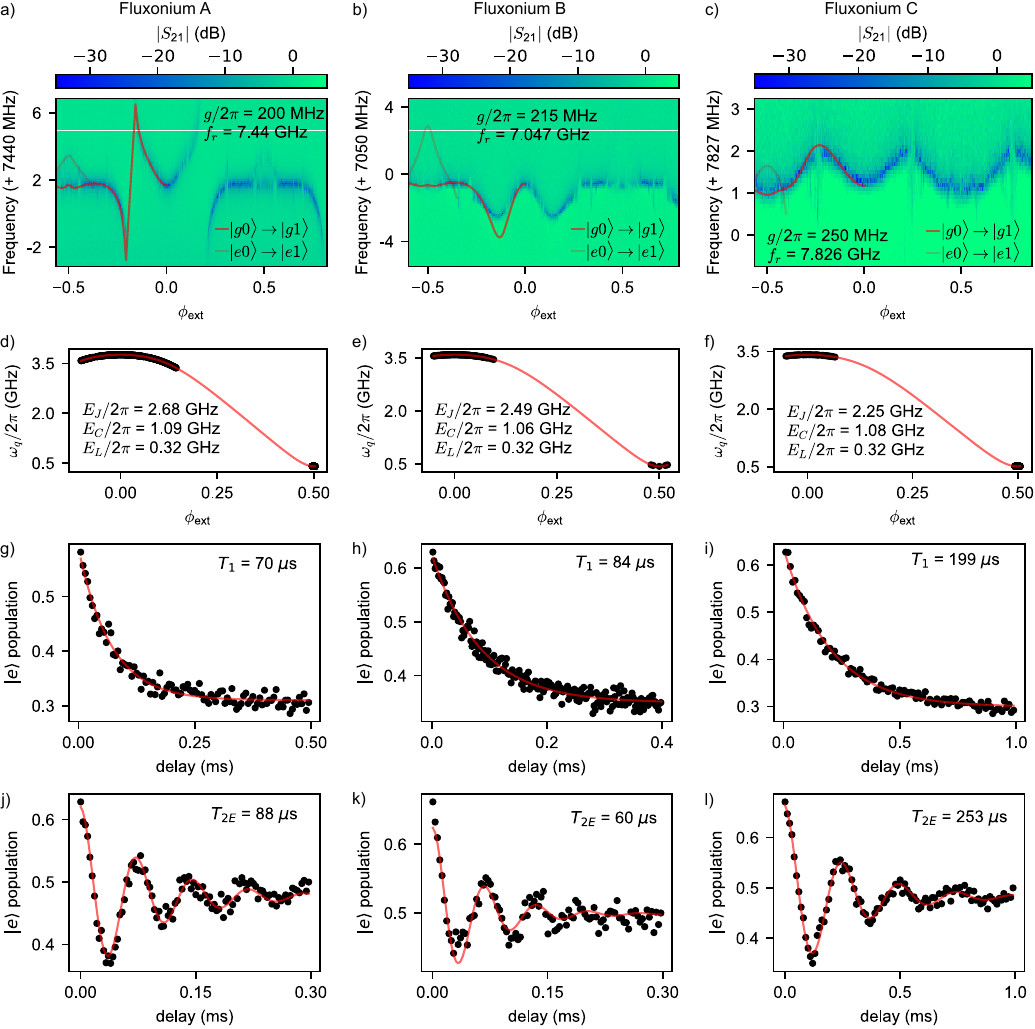}
        \caption{\label{device_characterization} 
            (a) ((b),(c)) Resonator flux spectroscopy fit for Device A (B, C). (d) ((e),(f)) Qubit flux spectroscopy fit for Device A (B, C). (g) ((h),(i)) Device A (B, C) qubit relaxation measurement. (j) ((k),(l)) Device A (B, C) qubit decoherence measurement with one echo pulse.}
    \end{figure*}

    We took resonator spectroscopy while sweeping $\phi_\mathrm{ext}$ to extract the frequency of the \qstate{g0} to \qstate{g1} transition as a function of $\phi_\mathrm{ext}$.
    Here, \qstate{in} represents the qubit state \qstate{i} and the resonator state \qstate{n}.
    The results of the spectroscopy for Device A, B, and C are shown in Figure~\ref{device_characterization}(a),(b),(c) respectively.
    We measured the fluxonium \qstate{g} to \qstate{e} frequency ($\omega_q$) around $\phi_\mathrm{ext} = 0$ and $\phi_\mathrm{ext} = 0.5$ and extracted the fluxonium parameters, $E_J$, $E_C$ and $E_L$ by fitting to the eigenvalues of the fluxonium Hamiltonian,

    \begin{equation} \label{flxm_ham}
    \hat{H}_f = 4E_C\hat{n}^2 + \frac{1}{2}E_L\hat{\phi}^2 - E_J\cos(\hat{\phi}-2\pi\phi_\mathrm{ext})
    \end{equation}

    Figure~\ref{device_characterization}(d),(e),(f) show these fits for Device A, B and C respectively.
    We extracted the bare resonator frequency ($f_r=\omega_r/2\pi$) and the coupling strength $g$ by fitting the resonator spectroscopy to the static fluxonium-resonator Hamiltonian,

    \begin{equation} \label{flxm_res_ham}
    \hat{H}_\mathrm{stat} = \hat{H}_f
    +\omega_r \hat{a}^\dag_r \hat{a}_r - i g\hat{n}(\hat{a}_r-\hat{a}^\dag_r)
    \end{equation}
    
    The fits to the resonator spectroscopy are included in Figure~\ref{device_characterization}(a),(b),(c).
    Near $\phi_\mathrm{ext} = 0.5$, as $f_q = 2 \pi\omega_q$ approaches 500 MHz, the population in \qstate{e} at thermal equilibrium increases, which result in the \qstate{e0} to \qstate{e1} transition also becoming visible in spectroscopy.
    
    To characterize the loss in the devices, we measured the relaxation time $T_1$ and the decoherence time with one echo pulse ($T_{2E}$) at $\phi_\mathrm{ext}$ = 0.5.
    The measured $T_1$ ($T_{2E}$) of Device A, B and C are shown in Figure~\ref{device_characterization}(g),(h),(i) ((j),(k),(l)) respectively.
    Both $T_1$ and $T_{2E}$ of the devices are on the order of 100 $\mu$s.
    We estimate that the main limit on qubit dephasing is due to photon shot noise from thermal photons in the resonator.
    
    \section{Spectroscopy of the fluxonium near predicted TLS}

\begin{figure}
    \centering
        \includegraphics{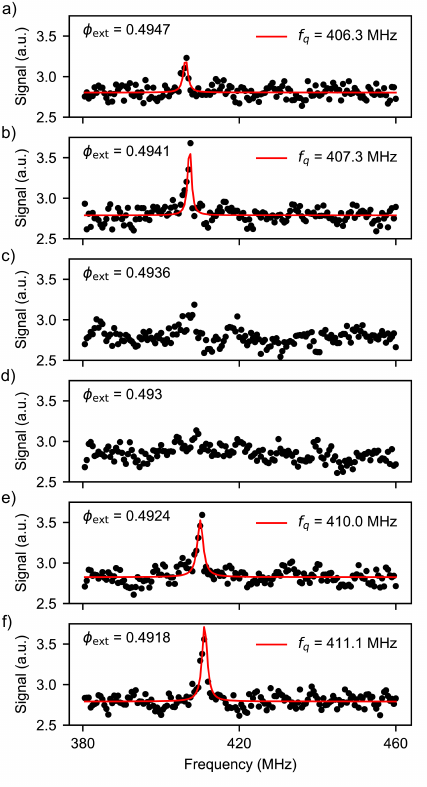}
        \caption{\label{tls_in_qubit_spec} 
            Qubit spectroscopy at different $\phi_\mathrm{ext}$ of Device A near the TLS frequency predicted from modeling (411 MHz).}
    \end{figure}

    As an additional check on the presence of TLS in our device, we may use qubit spectroscopy at differing frequencies to search for possible coupling to a TLS mode \cite{KlimovTLSspec}. 
    As the qubit transition frequency crosses the TLS frequency, an avoided crossing should form in the qubit spectrum with the width of the frequency splitting determined by the strength of the coupling to the TLS. 
    In order to deduce the frequency of the TLS, we performed qubit spectroscopy as a function of $\phi_\mathrm{ext}$ on Device A.
    While the linewidth of the qubit signal ($\sim$ 1.5 MHz) limits us from resolving splittings in the 1 MHz range, we do find that the qubit signal disappears near the TLS frequency predicted by our model (411 MHz) as shown in Figure~\ref{tls_in_qubit_spec}. 
    The measurements were taken over a small range of flux of 0.0029 $\phi_\mathrm{ext}$ which results in minimal variation of $\chi$ (1.1 kHz) and resonator frequency (20 kHz). 
    Hence, such a marked change in qubit frequency is not to be expected from changes in readout parameters only and this observation provides a strong indication of the presence of a TLS between 407 MHz and 410 MHz, close to the frequency expected from our model.
    
    \section{Resonance behavior near half flux}
    
\begin{figure}
    \centering
        \includegraphics{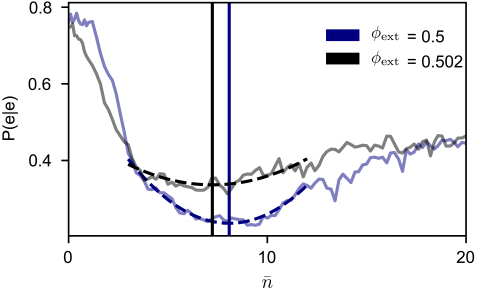}
        \caption{\label{fig:fluxcomparison} 
        Observed QND survival probability (dots) of the \qstate{e} state at half flux and at a closely detuned flux value of $\phi_\mathrm{ext} = 0.502$. To extract the resonance value $n^*$, indicated for each flux with a vertical line, we fit the local minima of each curve with a quadratic (dashed lines). This data was taken at a different date than in Fig.~2 of the main text.}
    \end{figure}

    As an additional check on the validity of the proposed TLS-fluxonium hybridization mechanism, we compare the $P(e|e)$ QND response of Device A at two closely spaced external fluxes of $\phi_{\mathrm{ext}} = 0.5$ and $\phi_{\mathrm{ext}} = 0.502$. For a fixed TLS at 411 MHz as predicted by the model, the small increase in qubit frequency from \abthree{401.5 MHz at} $\phi_{\mathrm{ext}} = 0.5$ to \abthree{402 MHz at} $\phi_{\mathrm{ext}} = 0.502$ that brings the qubit frequency closer to the TLS should result in the resonance occurring at lower photon number. The $P(g|e)$ resonance location $\bar{n}^*$ shifts as \begin{align}
        \delta \bar{n}^*_{\mathrm{pred.}} = -\frac{\delta \omega_{ge}}{2\chi}
    \end{align}
    due entirely to the flux-induced frequency shift $\delta \omega_{ge}$ of the fluxonium $g-e$ transition. We extract the observed resonance shift (see Figure \ref{fig:fluxcomparison}) as $\delta \bar{n}^*_{\mathrm{obs.}} \approx -0.85$ by curve-fitting the experimental data at both fluxes, and find that it is close in magnitude to its model-predicted value $\delta \bar{n}^*_{\mathrm{pred.}} \approx -0.28$ for Device A.
    
    \section{Higher energy excited-state readout}

    We directly drive the \qstate{e} to \qstate{f}, \qstate{g} to \qstate{h} and \qstate{e} to \qstate{i} transitions before a measurement pulse to observe the readout response for the \qstate{f}, \qstate{h} and \qstate{i} states.
    For all three devices, the distribution of the resonator frequencies corresponding to \qstate{g}, \qstate{e}, \qstate{f}, \qstate{h}, and \qstate{i} states spans a range of $\sim$ 10 $\kappa$, which made it difficult to use a single readout frequency that can discriminate between all the states simultaneously.
    We therefore optimized our readout frequency, amplitude, and length to best distinguish between the computational and non-computational states of the qubit while minimizing non-QND effects of our readout pulse.
    We find that, for Device A and C, the \qstate{f}, \qstate{h} and \qstate{i} are distinguishable from the \qstate{g} and \qstate{e} states but indistinguishable from each other.
    The histograms for these states are shown in Figure~\ref{upper_levels}.
    Our attempts to resolve the fluxonium states outside the qubit subspace in Device B by increasing the readout power were stymied by the onset of non-QND effects.
    Hence, for this device, we chose readout parameters that only resolve the \qstate{g} and \qstate{e} states.
    The measurement parameters used for the resonator responses shown in Figure~\ref{upper_levels} are the parameters for the final readout of Device A and C listed in Table~\ref{Supplemental Table 1}.

    \section{Relaxation times of fluxonium states}
    
    The relaxation times between states \qstate{e} and \qstate{f}, \qstate{g} and \qstate{h} and \qstate{e} and \qstate{i} for each device are listed in Table \ref{Supplemental Table 3}. 
    The delay between the end of a measurement run and the beginning of the next one is also included for comparison.
    \abthree{All listed $T_1$ values were obtained at $\phi_\mathrm{ext} = 0.5$.
    For device A, $T_1^{ge}$ at $\phi_\mathrm{ext} = 0.48$ and 0.52 was 152.4 $\mu$s.
    At these flux points, due to multiple decay routes, $T_1$ measurements for the higher states yielded multi-exponential decays which were difficult to fit.}
    
    \setlength{\tabcolsep}{7pt}
    \begin{table} [h!]
    \centering
    \caption{Measured state relaxation times and repetition delays (all quantities are in $\mu$s) \label{Supplemental Table 3}}
    \begin{tabular}{c  c  c  c  c c} 
     Device & $T_1^{ge}$ & $T_1^{ef}$ & $T_1^{gh}$ & $T_1^{ei}$ & Delay \\
     \hline
     A  & 70 & 25 & 63 & 10 & 500 \\
     B  & 84 & 14 & 65 & 17 & 500 \\
     C  & 200 & 23 & 227 & 51 & 1000 \\
     
    \end{tabular}
    \end{table}

    \section{Photon number calibration} 

\begin{figure}
    \centering
        \includegraphics{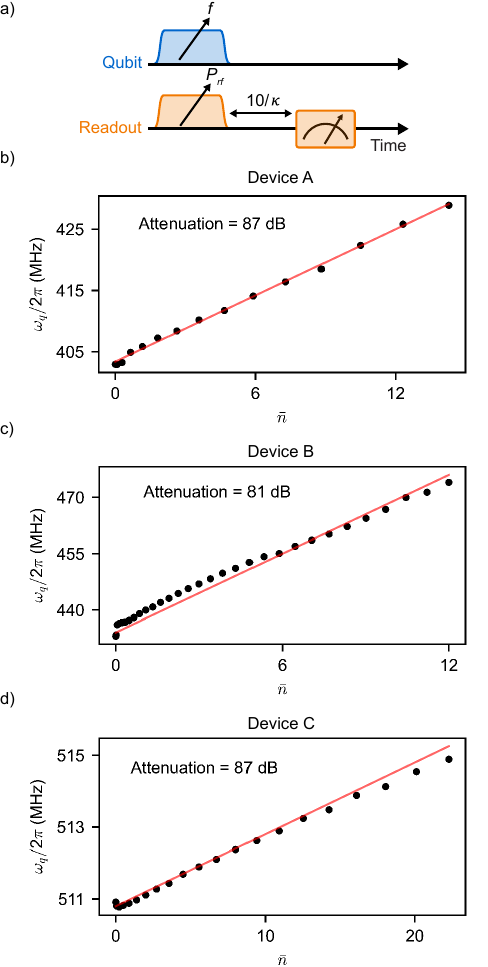}
        \caption{\label{n_bar_cal} 
            (a) The pulse sequence used to measure the ac-Stark shift of the qubit due to photon population in the resonator. (b),(c),(d) Fit of the Stark-shifted qubit frequency for Device A, B, and C.}
    \end{figure}

\begin{figure}
        \includegraphics{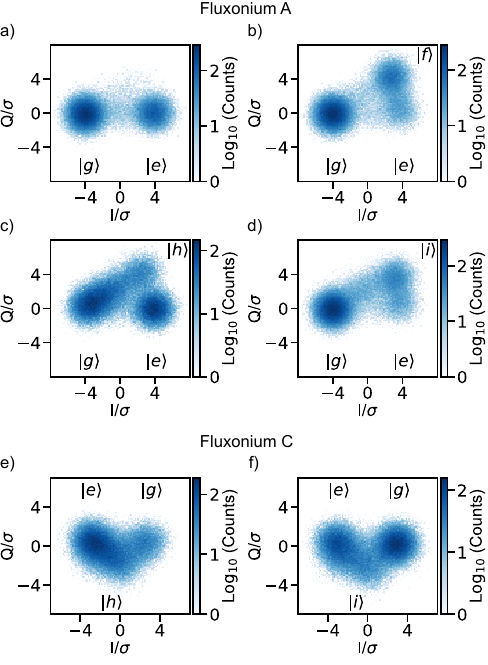}
        \caption{\label{upper_levels} 
            Device A: Resonator response after (a) $\hat{\mathrm{I}}$, (b) $\hat{\mathrm{X}}_\mathrm{ef}$, (c) $\hat{\mathrm{X}}_
            \mathrm{gh}$, (d) $\hat{\mathrm{X}}_\mathrm{ei}$. Device C: Resonator response after (e) $\hat{\mathrm{X}}_
            \mathrm{gh}$, (f) $\hat{\mathrm{X}}_\mathrm{ei}$.}
    \end{figure}

    We used the ac-Stark shift of the qubit frequency to calibrate the number of photons introduced by the proxy readout drive \cite{Gambetta_Qubit-photon}.
    The pulse sequence for the experiment is shown in Figure~\ref{n_bar_cal}(a).
    We simultaneously applied a saturation pulse on the qubit and a proxy readout tone on the resonator. 
    We then waited for a time 10/$\kappa$ let the photons evacuate the resonator and perform a readout.
    We repeated this experiment while sweeping the frequency of the qubit probe tone to measure the ac-Stark shift at a given proxy readout probe power ($P_{rf}$). 
    We sweep the frequency ($f$) of the qubit tone to perform spectroscopy while we sweep the probe power ($P_{rf}$) of the proxy readout tone to observe how the qubit spectrum changes with $P_{rf}$.
    The qubit frequency depends on the photon number through

    \begin{equation} \label{StarkShift}
    \begin{split}
    \omega_q(\bar{n}) = \omega_q(0)+2\chi\bar{n}
    \end{split}
    \end{equation}

    \noindent where $\bar{n}$ = $\bar{n}_+$ + $\bar{n}_-$ and $\bar{n}_+$ ($\bar{n}_-$) is the number of photons introduced by the proxy readout tone when the qubit is in \qstate{g} (\qstate{e}).
    To convert between the experimentally controlled variable $P_{rf}$ and $\bar{n}_\pm$, we follow \cite{Gambetta_Qubit-photon}, 

    \begin{equation} \label{Power}
    \begin{split}
    \bar{n}_\pm = \alpha \frac{\kappa/2}{\hbar\omega_{rf}((\kappa/2)^2+(\Delta \pm \chi  \pm K_\pm(\bar{n}_\pm)})^2)P_{rf},
    \end{split}
    \end{equation}    
    
    \noindent where $\alpha$ is a scaling factor that takes into account the attenuation seen by the proxy readout tone before reaching the sample.
    The frequency of the proxy readout tone is $\omega_{rf}$ and $\Delta = \omega_{rf} - \omega_r$\abthree{, where $\omega_r$ is the bare resonator frequency}.
    Coupling to a qubit introduces a small non-linearity to the resonator. 
    This photon number-dependent correction is represented by the Kerr coefficient, $K_\pm(\bar{n}_\pm)$.
    For all of the QNDness data reported in the main and the supplemental text, we set $\Delta=0$. 
    At each $P(f_0|i_0)$ point, $\bar{n} = \bar{n}_+ + 0$ when $i_0=g$ and $\bar{n} = 0 + \bar{n}_-$ when $i_0=e$.
    
    We independently measure the quantities $\omega_r$, $\kappa$, and $\chi$ from the spectroscopy of the resonator, which allows us to calculate $K_\pm$ from the system Hamiltonian.
    We control the frequency of the proxy readout tone to set $\Delta$ to any desired value.
    Combining these known quantities, we define constant quantities

    \begin{equation} \label{C}
    \begin{split}
    C_\pm = \frac{\kappa/2}{\hbar\omega_{rf}((\kappa/2)^2+(\Delta \pm \chi \pm K_\pm(\bar{n}_\pm)})^2)
    \end{split}
    \end{equation}  
    
    \noindent We can now combine equations (\ref{StarkShift}), (\ref{Power}) and (\ref{C}) to relate the qubit frequency to $P_{rf}$,

    \begin{equation} \label{StarkShiftPower}
    \begin{split}
    \omega_q(P_{rf}) = \omega_q(0)+2\chi\alpha(C_++C_-)P_{rf}
    \end{split}
    \end{equation}
    
    We performed the above-mentioned ac-Stark shift experiment at multiple different $P_{rf}$ to obtain the shifted qubit frequency as shown in Figure~\ref{n_bar_cal}.
    We then fit our ac-Stark shift data to equation~(\ref{StarkShiftPower}) with $\alpha$ as the only fit parameter.
    Once we know $\alpha$, we can use equations (\ref{Power}) to map any $P_{rf}$ value to $\bar{n}_\pm$.
    The data and fit for Device A, B and C are shown in Figure~\ref{n_bar_cal}(a),(b),(c) respectively. 
    The scaling factors are reported in the logarithmic scale.
    The input line has a total attenuation of 70 dB.
    We expect an additional attenuation of $\approx$ 10 dB from the microwave cables. 
    We attribute an additional attenuation of $\approx$ 6 dB for Device A and C to an impedance mismatch between the printed circuit board and a connector.
    Device B, which was measured using a different PCB in a different thermal cycle of the dilution refrigerator, did not face the same problem.
    
    \section{Readout pulses and state assignment}

\begin{figure}
    \centering
        \includegraphics{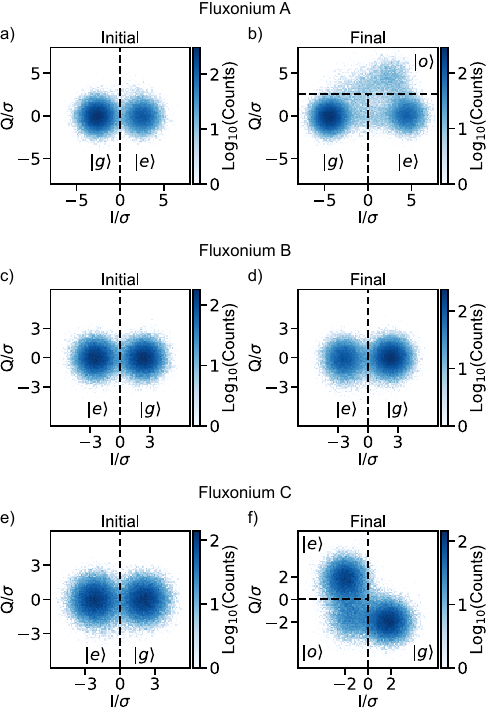}
        \caption{\label{state_assignment} 
            Device A: (a) State assignment criteria for initial measurement. Responses to the left (right) of the vertical dashed line are assigned as \qstate{g} (\qstate{e}). (b) State assignment for the final measurement. Responses above the horizontal dashed line are assigned as \qstate{o}. Of the remaining measurements, the ones to the left (right) of the vertical dashed line are assigned as \qstate{g} (\qstate{e}). Fifteen photons are introduced into the resonator between the initial and final measurements. Device B: (b,c) State assignment for the initial and final measurement. The vertical line represents the threshold for determining the qubit state. Ten photons are introduced into the resonator between the initial and final measurements. Device C: (e) State assignment for initial measurement. Responses to the right (left) of the vertical dashed line are assigned as \qstate{g} (\qstate{e}). (f) State assignment for the final measurement. Responses to the right of the vertical dashed line are assigned as \qstate{g}. Of the remaining measurements, the ones above (below) the horizontal dashed line are assigned as \qstate{e} (\qstate{o}). One hundred photons are introduced into the resonator between the initial and final measurements.}
    \end{figure}

    For each of the $P(f_0|i_0)$ datasets we report, we repeated the experiment described in the main text Figure 1(b) $10^5$ times.
    We used the $10^5$ responses of the initial and final measurements to create histograms with 101 bins.
    We fitted the histograms to Gaussian distributions to find the center and full width at half maximum.
    The initial measurements yield two Gaussian distributions, one for \qstate{g} and another for \qstate{e}.
    We rotated the data about the origin so that the Gaussians are aligned along the in-phase component (I = Re[$S_{21}$]) axis.
    We set the midpoint between of the two Gaussians as the threshold for state discrimination.
    The readout histograms for the initial measurements for Device A, B and C are shown in Figure~\ref{state_assignment}(a),(c),(e) respectively.
    The threshold for state discrimination is marked by a vertical black dashed line.

    The parameters of the final measurement of Device B are identical to its initial measurement and we use the same method of state discrimination (Figure~\ref{state_assignment}(d)).
    We found it possible to obtain up to three Gaussian distributions (\qstate{g}, \qstate{e} and \qstate{o}) for the final measurements of Device A and C. 
    For Device A, we rotated the data so that the Gaussians corresponding to the qubit states are aligned along the I axis and set the threshold at the midpoint as shown by the vertical black dashed line in Figure~\ref{state_assignment}(b). 

    We are able to discriminate the Gaussian corresponding to \qstate{o} from the \qstate{g} and \qstate{e} Gasussians along the quadrature component (Q = Im[$S_{21}$]) axis.
    We set the threshold at the midpoint between the centers of the \qstate{g} and \qstate{e} Gaussians and the \qstate{o} Gaussian along the Q axis as shown by the horizontal dashed line in Figure~\ref{state_assignment}(b).
    For Device C, we rotated the data so that the Gaussians corresponding to \qstate{g} and \qstate{o} are aligned along the I axis while the Gaussians corresponding to \qstate{e} and \qstate{o} are aligned along the Q axis (Figure~\ref{state_assignment}(f)).
    We set the threshold for discriminating \qstate{g} and \qstate{e} at the midpoint between the center of their Gaussians along the I axis as shown by the vertical black dashed line in Figure~\ref{state_assignment}(f).
    We set the threshold for discriminating \qstate{o} and \qstate{e} at the midpoint between the center of their Gaussians along the Q axis as shown by the horizontal black dashed line in Figure~\ref{state_assignment}(f).
    The parameters used in the measurement of each fluxonium are shown in Supplemental Table~\ref{Supplemental Table 1}.

    \section{Assignment errors}
    
    For two Gaussians with centers $I_0$ and $I_1$ (corresponding to states \qstate{g} and \qstate{e}) with  standard deviation $\sigma$ and the threshold for state discrimination at the midpoint of the two Gaussians, ($I_0$ + $I_1$)/2, assuming an infinite qubit lifetime, the probabilities of incorrect assignment of states are given by \cite{Gambetta_Protocols, Krantz_QuantumEngineer}, 

    \begin{equation} \label{GaussianError1}
    \begin{split}
    P_{err}(\mathrm{e}|\mathrm{g}) = \frac{1}{\sigma\sqrt{2\pi}}\int_{(I_0 + I_1)/2}^\infty e^{-\frac{(I-I_0)^2}{2\sigma^2}}dI
    \end{split}
    \end{equation}

    \begin{equation} \label{GaussianError2}
    \begin{split}
    P_{err}(\mathrm{g}|\mathrm{e}) = \frac{1}{\sigma\sqrt{2\pi}}\int_{(I_0 + I_1)/2}^\infty e^{-\frac{(I-I_1)^2}{2\sigma^2}}dI
    \end{split}
    \end{equation}

    \noindent Solving the integrals in (\ref{GaussianError1}) and (\ref{GaussianError2}),
    
    \begin{equation} \label{ErrorProb}
    \begin{split}
    P_{err}(\mathrm{e}|\mathrm{g}) = P_{err}(\mathrm{g}|\mathrm{e}) = \frac{1}{2}(1-\mathrm{Erf}[\sqrt{\frac{\mathrm{SNR}}{8}}]) 
    \end{split}
    \end{equation}
    \noindent where SNR = $(\frac{I_0-I_1}{\sigma})^2$.
    We obtain the SNR and error probability between any two states from the Gaussian fits to the readout histograms. 
    In a two-state measurement, the counts for each state $g_i$ and $e_i$,

    \[
    C_i = \begin{pmatrix} g_i \\ e_i \end{pmatrix} ,
    \]

    \noindent are transformed by the error matrix,

    \[
    E_i = \begin{pmatrix} P_{{err}_i}'(\mathrm{e}|\mathrm{g}) & P_{{err}_i}(\mathrm{g}|\mathrm{e}) \\ P_{{err}_i}(\mathrm{e}|\mathrm{g}) & P_{{err}_i}'(\mathrm{g}|\mathrm{e}) \end{pmatrix} ,
    \]
    \noindent where $P_{{err}_i}'(f_0|i_0) = 1-P_{{err}_i}(f_0|i_0)$, to the measured counts,
    \begin{equation} \label{InitError}
    \begin{split}
    C_{m_i} = E_iC_i
    \end{split}
    \end{equation}
    
    \noindent One can correct for this error by inverting $E_i$ and applying it to $C_{m_i}$,

    \begin{equation} \label{InitErrorCorrection}
    \begin{split}
    C_i = E_i^{-1}C_{m_i}
    \end{split}
    \end{equation}

    \noindent Similarly, for three-state measurements, the counts for each state $g_f$, $e_f$ and $o_f$,

    \[
    C_f = \begin{pmatrix} g_f \\ e_f \\ o_f \end{pmatrix} ,
    \]

    \noindent are transformed by the error matrix,

    \[
    E_f = \begin{pmatrix}
    P'_{{err}_f}(f_0|\mathrm{g}) & P_{{err}_f}(\mathrm{g}|\mathrm{e}) & P_{{err}_f}(\mathrm{g}|\mathrm{o}) \\ P_{{err}_f}(\mathrm{e}|\mathrm{g}) & P'_{{err}_f}(f_0|\mathrm{e}) & P_{{err}_f}(\mathrm{o}|\mathrm{e}) \\ P_{{err}_f}(\mathrm{o}|\mathrm{g}) & P_{{err}_f}(\mathrm{o}|\mathrm{e}) & P'_{{err}_f}(f_0|\mathrm{o})\end{pmatrix} ,
    \]

    \noindent where, $P'_{{err}_f}(f_0|i_0) = 1 - \sum_{i\neq i_0}P_{{err}_f}(i|i_0)$ with $i$ = [g, e, o], to the measured counts,
    
    \begin{equation} \label{FinError}
    \begin{split}
    C_{m_f} = E_fC_f,
    \end{split}
    \end{equation} 

    \noindent and one can correct for this error by inverting $E_f$ and applying it to $C_{m_f}$,

    \begin{equation} \label{FinErrorCorrection}
    \begin{split}
    C_f = E_f^{-1}C_{m_f}
    \end{split}
    \end{equation}
    
    \begingroup
    \setlength{\tabcolsep}{7pt}
    \begin{table*}
    \centering
    \caption{Readout parameters \label{Supplemental Table 1}}
    \begin{tabular}{c  c  c  c  c  c  c  c} 
     Device & Readout Type & $\bar{n}^r_+$ & $\bar{n}^r_-$ & Length ($\mu$s) & $P_{err}(e|g)$ & $P_{err}(o|g)$ & $P_{err}(o|e)$ \\
     \hline
     A  & initial  & 2.8 & 0.1 & 5 & 5$\times 10^{-3}$ & - & - \\
     A  & final  & 4.1 & 0.1 & 10 & 2$\times 10^{-6}$ & 6$\times 10^{-3}$ & 6$\times 10^{-3}$ \\
     B  & initial  & 2.4 & 0.0 & 10 & 7 $\times 10^{-3}$ & - & - \\
     B  & final  & 2.4 & 0.0 & 10 & 7 $\times 10^{-3}$ & - & - \\
     C  & initial  & 0.3 & 0.2 & 10 & 2 $\times 10^{-2}$ & - & - \\
     C  & final  & 0.3 & 0.2 & 20 & 2 $\times 10^{-2}$ & 2 $\times 10^{-2}$ & 2 $\times 10^{-2}$ \\
     
    \end{tabular}
    \end{table*}
    \endgroup

    \begin{figure}
    \centering
        \includegraphics{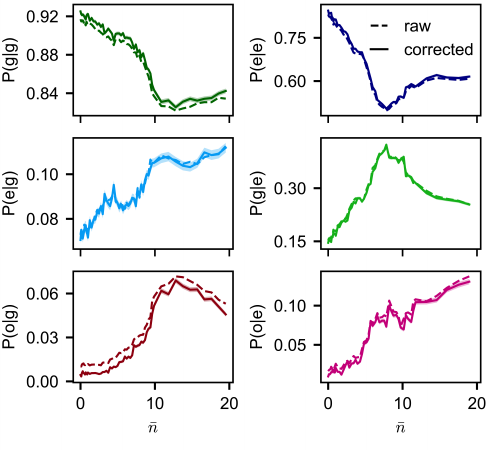}
        \caption{\label{err_analysis} 
            The raw and processed data for the probability of the Device A staying in \qstate{g} (\qstate{e}) or transitioning to \qstate{e} (\qstate{g}) or \qstate{o} as a function of $\bar{n}$ taken at $\phi_\mathrm{ext} = 0.5$. The dashed lines show the raw data. The solid lines show the mean value of the error corrected bootstrapped samples and the filling around the solid lines show their standard deviation.  }
    \end{figure}
    Additional error is introduced in the state assignment due to spurious qubit jumps during the measurement. 
    One can correct for this error if the relaxation and excitation rates of all the states involved are well known.
    In our case, there are multiple possible decay channels for \qstate{o} and the relaxation and excitation rates to each state from \qstate{o} cannot be determined.
    Hence, we refrain from correcting errors related to spurious qubit jumps.

    We used bootstrapping to estimate the variation in our data and determine the errorbars.
    For each dataset, we create $10^3$ samples of size $2\times10^4$.
    We used (\ref{InitErrorCorrection}) and (\ref{FinErrorCorrection}) to correct for the error in each sample.
    We present the mean probabilities found from the samples in our figures.
    We find that the errorbars, given by the standard deviation in the probabilities found from each sample, are smaller than the markers used in the main text.
    The raw and processed probabilities for the data in Figure 2(a),(b) of the main text is shown in Figure~\ref{err_analysis}.
    The errors in the measurements of each device are detailed in Table~\ref{Supplemental Table 1}.

    \section{Changes in non-QND effects over time and thermal cycle}
    \begin{figure*}
    \centering
        \includegraphics{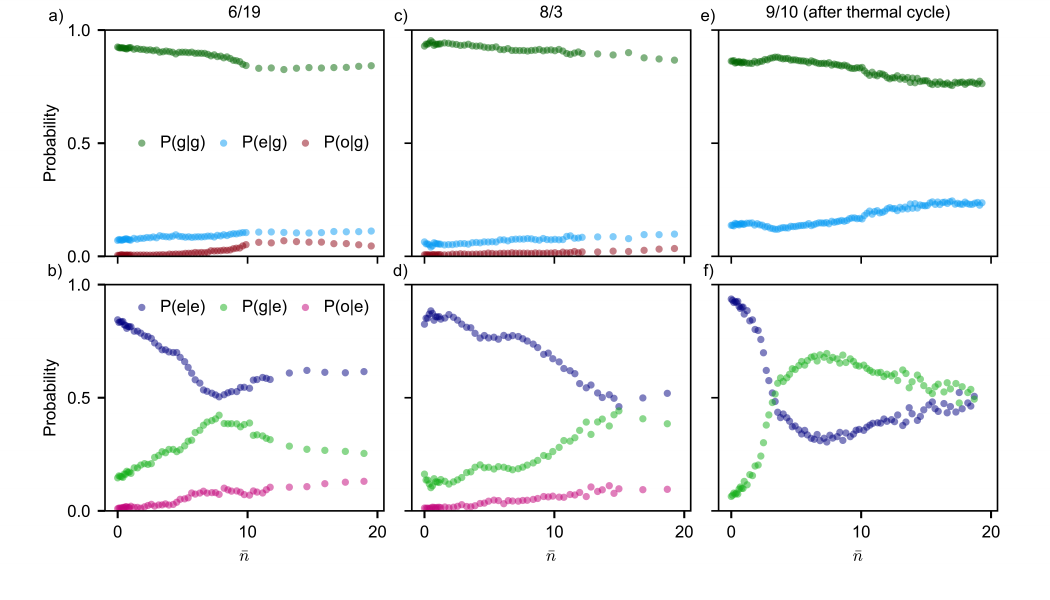}
        \caption{\label{flxmA_diff_days} 
            The probability of the Device A staying in \qstate{g} (\qstate{e}) or transitioning to \qstate{e} (\qstate{g}) or \qstate{o} as a function of $\bar{n}$ over different days. The subheadings show the date on which the data was taken.}
    \end{figure*}
    We repeated the experiment described in the main text on Device A over the span of multiple months and cooldowns.
    The data in the main text was taken on 6/19/24 and is shown in Fig.~\ref{flxmA_diff_days}(a),(b). We repeated the experiment in the same cooldown on 8/3/24 as shown in Figure~\ref{flxmA_diff_days}(c),(d).
    We observe that the behavior when the initial state is \qstate{g} is similar in both sets of measurement with $P(g|g)$ dropping to $\approx 0.85$ at $\bar{n} = 20$.
    In the case when the initial state is \qstate{e}, we see that the position of the resonance feature with respect to $\bar{n}$ changes from $\bar{n} \approx 7$ to $\bar{n} \approx 12$. 
    The fluxonium frequency and Hamiltonian parameters remained the same over this time span, which suggests that the TLS environment changed between the two datasets.
    We also measured the same device after a full thermal cycle of the sample on 9/10/24 as shown in Figure~\ref{flxmA_diff_days}(e),(f).
    The Hamiltonian parameters of the device again remained the same during this cooldown.
    Due to changes in the $T_1$ of the \qstate{f}, \qstate{h} and \qstate{i} states, however, we could not discriminate the states outside the qubit manifold.
    Hence, we only report the probabilities of the final state being \qstate{g} or \qstate{e}.
    When the initial state is \qstate{g}, we observe slightly more non-QND behavior for $\bar{n} > 10$ with $P(g|g)$ dropping to $\approx 0.76$ at $\bar{n} = 20$.
    When the initial state is \qstate{e}, we find a much broader resonance feature occurring at $\bar{n} \approx 7$.
    Spectral fluctuations in TLS in a single cooldowns and over multiple cooldowns and the consequent changes in qubit relaxation rates have been previously observed \cite{Burnett_decoherence, Thorbeck_TLS}.
    Since we have not observed any changes in the Hamiltonian parameters of the fluxonium and resonator, we attribute the changes observed in these datasets to changes in TLS parameters.

    \section{Simulation methods}

    To quantitatively understand the dynamics of the readout protocol we perform time-domain Lindbladian simulations of the driven fluxonium-resonator system both with and without a coupled TLS \cite{BlaisCED2021}. Our starting point is the Hamiltonian
    \begin{equation} \label{total_ham}
    \begin{split}
    \hat{H}_{\epsilon,\nu}(t) = 4E_C\hat{n}^2 + \frac{1}{2}E_L\hat{\phi}^2 - E_J\cos(\hat{\phi}-2\pi\phi_\mathrm{ext})\\
    +\omega_r \hat{a}^\dagger_r \hat{a}_r - ig\hat{n}(\hat{a}_r-\hat{a}^\dagger_r)
    - i\epsilon\cos\Omega t \,(\hat a_r  - \hat a_r^\dagger)\\
    + \nu\left( \,\Delta_{TLS}\hat Z_{TLS} + g_{TLS} \,\hat X_{TLS}\,\hat n \right) ,
    \end{split}
    \end{equation}
    with $\nu = 0$ (no TLS present) or $\nu = 1$ (one TLS present at frequency $\Delta_{TLS}$). $\hat Z$ and $\hat X$ are the Pauli operators. This Hamiltonian models the resonator driven monochromatically at frequency $\Omega$ and drive strength $\epsilon$. Since we are interested in the behavior when the resonator population reaches a fixed $\bar n$ upon application of a time-dependent drive, we promote to Lindbladian dynamics and incorporate jump operators $\hat a$ and $\hat a^\dagger$, resulting in the Lindblad superoperator\label{total_lind}
    \begin{align}
    \mathcal{L}_{\epsilon,\nu}(t)\rho &= -i[\hat H_{\epsilon,\nu}(t), \rho] + \kappa D[\hat a]\rho
    \end{align}
    \begin{align}\label{eq:dissipator}
    D[\hat{\mathcal O}]\hat\rho &= \hat{\mathcal O}\hat \rho \hat{\mathcal O}^\dagger - \tfrac 1 2 \{\hat{\mathcal O}^\dagger \hat{\mathcal O}, \hat\rho\}
    \end{align}
    where $\kappa$ is the decay rate of the resonator.

    We perform our simulations by passing to the Floquet frequency lattice, eliminating the explicit time dependence of $\mathcal L(t)$ by tensoring the system with a synthetic 1-D lattice of Fourier modes which represents the drive \cite{Holthaus_floquet}.
    This replaces the driven Hamiltonian $\hat H_{\epsilon,\nu}(t)$ with a Floquet Hamiltonian $\hat F_{\epsilon, \nu}$. 
    This simplifies our Lindbladian $\mathcal{L}_{\epsilon,\nu}(t) \to \mathcal K_{\nu, \epsilon}$. 
    We also make the assumption that the resonator decays primarily into the drive line. 
    With this in mind, the appropriate modification to the dissipator in equation (\ref{eq:dissipator}) is
    \begin{align}
        D[\hat a] \to D[\hat a \,\hat b^\dagger]
    \end{align}
    where $b^\dagger$ is the right-translation operator on the frequency lattice. With this change, the Lindbladian $\mathcal K$ is approximately energy-conserving. 
    
    We begin our simulations at fixed $\epsilon$ and fluxonium state \qstate{\psi} with sparse diagonalization of $F$ around an energy window $E_0 = \tr \hat F\,\hat \rho_{\psi,0}$. 
    The TLS (if present) is always initialized in its ground state for the simulations in the main text, but in Supplementary section XVI, we also show the effects of finite temperature by initializing the TLS into a thermal state. 
    In either case, this yields a basis of Floquet quasi-eigenstates within which we perform Lindblad real-time evolution. 
    Starting in an initial state $\hat \rho_{\psi,0} = \hat \psi \otimes|0\rangle\langle 0|$, where $\hat \psi$ is some state of the fluxonium and TLS and $|0\rangle\langle 0|$ is the resonator ground state, we expect that time evolution under $\mathcal K$ for times $\gg \kappa^{-1}$ will drive the system to a fixed point $\hat \rho_{\psi,\epsilon}$. In this final state, we define $\bar n(\epsilon) = \tr \hat n \hat \rho_{\psi,\epsilon}$ and the qubit measurement probabilities $p_{\phi}(\epsilon) = \tr \hat \Pi_\phi \,\hat \rho_{\psi,\epsilon}$ using the projections $\hat \Pi_\phi = |\phi\rangle\langle\phi|$.
    We perform this evolution for $10\times 2\pi/\kappa$ under $\mathcal K$ where $\bar n(\epsilon)$ has converged, and record $p_\phi$ for each bare fluxonium eigenstate \qstate{\phi}. 
    Repeating this for each initial fluxonium state over a range of $\epsilon$ yields transition probability curves over a range of $\bar n$, and we do so for both TLS options $\nu = 0,1$. For further details on this simulation methodology, see \cite{paper_TK}.

    The fluxonium, resonator, and TLS parameters for Device A are provided in the main text. 
    For Device B, the fitted TLS frequency is 430.5 MHz and fluxonium-TLS coupling is 1.5 MHz. 
    The fitted TLS frequency is 524.5 MHz and fluxonium-TLS coupling is 2.25 MHz for Device C. 
    \abthree{We include 10 fluxonium levels, 65 resonator Fock levels and 13 Floquet sidebands in our simulations.}

    \section{Device A simulation with no flux correction}

\begin{figure}
    \centering
        \includegraphics{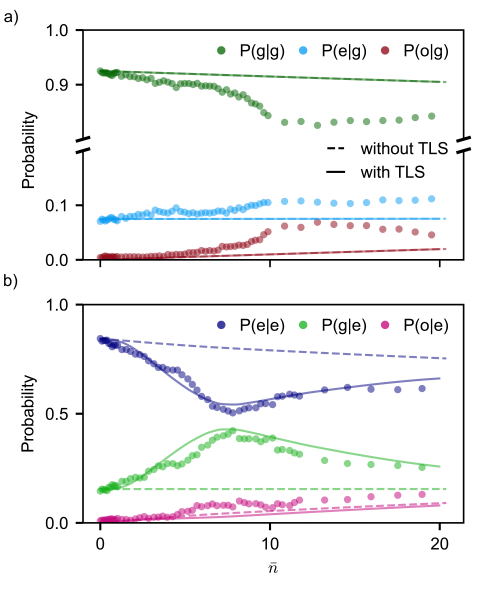}
        \caption{\label{fig:no_flux_correction} 
            QNDness of Device A at half-flux from Fig. 2 of the main text with simulations performed at $\phi_\mathrm{ext}$ = 0.5. Simulations performed exactly at half-flux give slightly worse fits to the data for the $P(g|g)$ and $P(o|g)$ probabilities.}
    \end{figure}

    The simulations performed exactly at half-flux for Device A are shown in Figure~\ref{fig:no_flux_correction}. 
    Compared to Fig. 2 of the main text, we see a larger discrepancy between simulation and data for $P(g|g)$ and $P(o|g)$ probabilities.
    Small deviations from half-flux, that are experimentally indiscernible, introduce non-zero matrix elements between $|g\rangle$ and $|i\rangle$ and lead to transitions between these states.

    \section{Floquet branch analysis}
\begin{figure}
    \centering
        \includegraphics{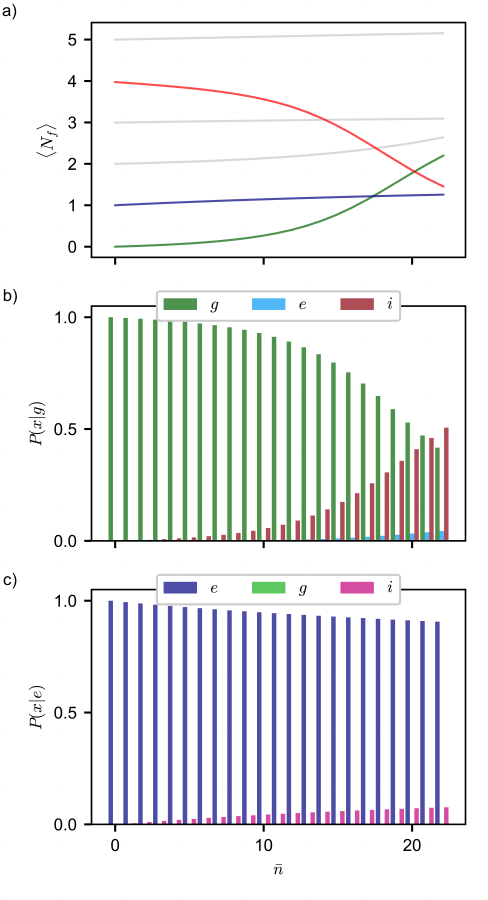}
        \caption{\label{fig:branchanalysis} 
        Branch analysis of the fluxonium at external flux $\phi = 0.52$ and drive strength $\epsilon = 0$ with no TLS present. (a) The change in the average fluxonium population $\langle N_f\rangle$ as the resonator photon number $\bar{n}$ increases is shown for the qubit initialized in the first 6 fluxonium states. For each branch state \qstate{\phi_{t;n}} $\in B_{g}$ (b) or $B_{e}$ (c), the vertical bars indicate the corresponding fluxonium measurement probabilities $\langle \psi_n |\Pi_x|\psi_n\rangle$. The resonator photons cause exchange of fluxonium population between \qstate{g} and \qstate{e}, and \qstate{i}. This behavior matches the results of the Floquet-Lindblad dynamical simulation.}
    \end{figure}

    Readout-induced state transitions of superconducting qubits have previously been extensively investigated via branch analysis \cite{Dumas_TransmonIonization, Nesterov_MIST, singh_impact_2024}. This scheme analyzes the static properties of the joint Floquet resonator-qubit Hamiltonian $H$ at drive strength $\epsilon$ by partitioning its set of eigenvectors $\sigma(H)$ according to the corresponding levels of the fluxonium. 
    We compare our Floquet-Lindblad dynamics algorithm used in the main text to the Floquet branch analysis (as shown in Fig.~\ref{fig:branchanalysis}), of Device A at external flux $\phi = 0.52$ and drive strength $\epsilon = 0$. 
    We observe, that similar to our analysis, the Floquet branch analysis also predicts significant population transfer out of the qubit states \qstate{g} and \qstate{e} into the \qstate{i} state at photon numbers above $\bar{n}\approx5$. 
    
    \section{Landau-Zener physics of qubit-resonator avoided crossings}
    Based on our simulations and the Floquet branch analysis, we expect that there is hybridization between the qubit states and the \qstate{i} upon introduction of a few readout photons. 
    The energy levels thus have an avoided crossing at a critical photon number, $\bar{n}_\mathrm{crit}$ that may be traversed diabatically or adiabatically depending on the rate at which the resonator photon number (here acting as the control field that determines the adiabaticity of the crossing) is increased \cite{Khezri_MISTWithin,Sank_MISTBeyond,Dumas_TransmonIonization}.
    When we experimentally apply a drive to the resonator, the photon number in the resonator must go smoothly from zero to $\bar{n}$, where $\bar{n}$ is the steady-state photon number as determined by the drive and resonator decay rate.
    Thus once the $\bar{n}$ in the resonator is larger than $\bar{n}_\mathrm{crit}$, these Landau-Zener effects determine the qubit population.
    Depending on whether the resonator photon number moves adiabatically across the critical photon number at which the resonant condition between the resonator and the higher-energy level of the fluxonium is met, it may either remain in the same fluxonium state (for a diabatic transition) or change state (for an adiabatic transition).
    Recent measurements of Landau-Zener effects in the transmon have shown that pulse shaping can be used to determine whether the transition is adiabatic or diabatic \cite{wangExcitedStateDynamics}.
    In our experiment, we \abthree{use square pulses and} do not specifically shape our pulse to quickly traverse across the avoided crossing. We therefore expect that adiabatic transitions occur as we increase the photon number.
    The Landau-Zener probability of a diabatic transition is:
    \begin{equation}
        P_\mathrm{LZ} = e^{-\pi \Delta^2_\mathrm{ac}/2 v},
    \end{equation}
    \noindent where $\Delta_\mathrm{ac}$ is the size of avoided crossing between the hybridized states and $v$ is the velocity at which the crossing is traversed. 
    We estimate from our simulation that the energy gap between $|g,n_\mathrm{crit}\rangle$ and $|i,n_\mathrm{crit}\rangle$ is 37.7 MHz and the velocity is roughly $\bar{n}\kappa$ for our unshaped pulse, which gives $P_\mathrm{LZ}\approx0$, meaning that all transitions above $\bar{n}_\mathrm{crit}$ are adiabatic.
    The adiabatic traversal results in the loss of population in the initial qubit states and the loss of QND-ness that we observe in our experiment for all photon numbers larger than $\bar{n}_\mathrm{crit}$.

    \section{Effect of temperature on TLS modeling}

    Previous measurements of TLS's in superconducting circuits have observed differing thermalization behaviors including coupling to a two-level fluctuator bath \cite{deGraafFluctuators}, strong coupling to phonon baths \cite{OdehPhononicBandgap}, as well as extremely weak coupling to phonons \cite{LiuDiscreteChargeStates}.
    These differing observed behaviors make it difficult to set a temperature for the TLS in our model.  
    While we initialized the TLS in the ground state in the main text, we show here the differing behaviors that arise for different TLS temperatures. 
    For a TLS that mediates a $k$-photon $g-e$ transition, the Boltzmann occupation of the excited state is
    \begin{align}
        p_{\mathrm{th}}(k) &= \frac{1}{1 + e^{\beta\Delta_0 + k\beta \Omega }}\\
        &\approx\frac{1}{1 + 1.93 \times \exp(11.9\times k)}
    \end{align}
    where we have taken $1/\beta = k_B \times 30\,\mathrm{mK}$.
    
    A TLS reasonably well-thermalized to the dilution refrigerator temperature of 10 mK that mediates a $0$-photon resonance may thus expected to have a nontrivial thermal excited state population of up to $34.1\,\%$, while that population falls below $10^{-10}$ for all $k \geq 2$. 
    This manifests in the simulation as a corresponding decrease in the $P(g|g)$ QND probabilities, for $k = 0$ only, around the resonance point $\bar{n}^*$, as the \qstate{g, 1_{\mathrm{TLS}}}$\longrightarrow $\qstate{e, 0_{\mathrm{TLS}}} transition is roughly speaking symmetric with the \qstate{g, 1_{\mathrm{TLS}}}$\longleftarrow $\qstate{e, 0_{\mathrm{TLS}}} transition.
    
\begin{figure}
    \centering
        \includegraphics{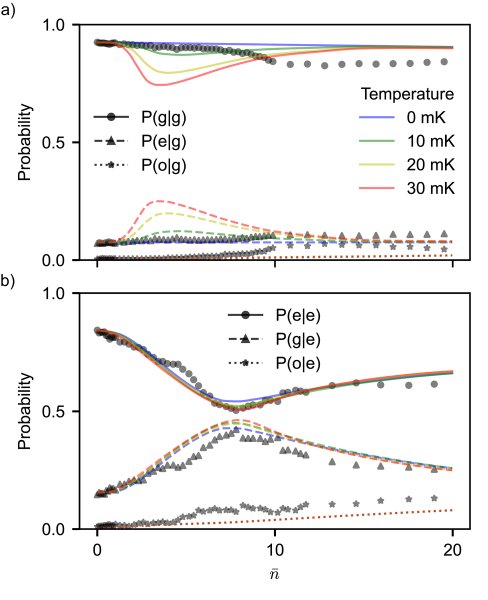}
        \caption{\label{tls_temp} 
        Results of modeling the QNDness behavior of Device A at half flux for a thermalized $0$-photon TLS over a range of temperatures $T$, with initial fluxonium state (a) \qstate{g} and (b) \qstate{e}. The lines show the simulation results and the markers show the data. The $T = 0$ model included in the main text is shown in blue.}
    \end{figure}

    We have quantified the QNDness effect of finite temperature via simulation for a range of temperatures $T$ between 0 and 30 $\textrm{mK}$ in Figure \ref{tls_temp}, where the TLS gap $\Delta_\mathrm{TLS}(T)$ and coupling $g_\mathrm{TLS}(T)$ are optimized for each $T$ to match the $P(e|e)$ QNDness data and are listed in Table \ref{Supplemental Table 2}. 
    For all finite temperatures, there is a visible discrepancy between the simulation, which displays a resonant drop in $P(g|g)$, and the experiment, which has no such resonance at that $\bar{n}^*$. 
    Future experiments on the bath that the TLS is thermalizing to will allow identification of the TLS temperature.
    
    \setlength{\tabcolsep}{7pt}
    \begin{table}[h!]
    \centering
    \caption{TLS parameters used in simulations with different TLS temperatures \label{Supplemental Table 2}}
    \begin{tabular}{c  c  c  c  c c} 
     Temperature (mK) & $\Delta_\mathrm{TLS}$ (MHz) & $g_\mathrm{TLS}$ (MHz) \\
     \hline
     0  & 411.0 & 1.30 \\
     10  & 410.5 & 1.45 \\
     20  & 410.0 & 1.60 \\
     20  & 409.5 & 1.75 \\
     
    \end{tabular}
    \end{table}

    \section{Spurious modes resonant with multi-photon transitions}
    The TLS frequencies were fitted so that resonance with the qubit occurred at first order in perturbation theory with respect to the couplings $g$, $g_{TLS}$, and $\epsilon$. 
    On the other hand, if a TLS gap of $\Delta$ fits a certain $\bar{n}$-resonance feature via a first-order (zero-photon) transition, a gap of $\Delta + 2m\Omega$ for integral $m > 0$ (a $2m$-photon transition) can result in the same resonance location and occurs at order $O(g\cdot g_{TLS}\cdot \epsilon^{2m})$. 
    Because of this multi-photon hybridization path, it is difficult to determine the effects of a spurious mode with transition corresponding to $m$-photons for different $m$ using the $P(f_0|i_0)$ vs. $\bar{n}$ QNDness data at a fixed flux point.
    
\begin{figure}
    \centering
        \includegraphics{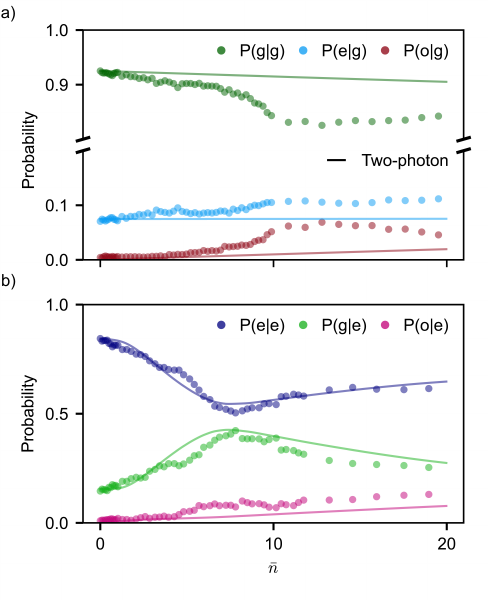}
        \caption{\label{multi-photon_flxmA} 
        Results of modeling the QNDness behavior of Device A at half flux for initial state (a) \qstate{g} and (b) \qstate{e} using a 2 photon transition with corresponding TLS frequency = 15.2991 GHz and coupling $g_\mathrm{TLS}=330\,\mathrm{MHz}$.}
    \end{figure}

    If we consider the $m = 2$ transition in particular, which would likely be close to the frequency of a Josephson junction array (JJA) mode, the resonance occurs at order $O(g\cdot g_{TLS}\cdot \epsilon^2)$ in perturbation theory, and obtaining the same hybridization phenomena as the $m = 0$ case requires a corresponding increase in $g_{TLS}$. 
    To quantify this, we perform the same fitting procedure with the spurious mode gap now constrained to be near $\Delta_2 = 2\Omega + \Delta_0$, where $\Delta_0 = 411\,\mathrm{MHz}$ was the best-fit zero-photon TLS. 
    The results are shown in Fig.~\ref{multi-photon_flxmA} in analogy to Fig.~2 in the main text. 
    We find that a much larger coupling of $g_{TLS} = 330\,\mathrm{MHz}$, and mode frequency of $\Delta_2 = 15.2901\,\mathrm{GHz}\approx 2\Omega + \Delta_0$, approximately reproduces the best-fit from the zero-photon case.
    
    Although coupling to this JJA array mode can be fit to the data, our spectroscopy and QNDness measurements suggest that materials defects are likely playing a more important role in determining non-QNDness in our devices. Moreover, our observations of the changing QNDness behavior within and across different cooldowns are incompatible with the expected behavior of JJA modes, which should not change with time or cooldown number as long as the Hamiltonian parameters remain the same.

\nocite{*}

\end{document}